\font\tenmsy=msbm10
\font\sevenmsy=msbm10 at 7pt
\font\fivemsy=msbm10 at 5pt
\newfam\msyfam 
\textfont\msyfam=\tenmsy
\scriptfont\msyfam=\sevenmsy
\scriptscriptfont\msyfam=\fivemsy
\def\blackB{\fam\msyfam\tenmsy}
\newcommand{\nc}{\newcommand}
\def\ZZ{{\blackB Z}}
\def\QQ{{\blackB Q}}

\def\NN{{\blackB N}}

\def\KK{{\blackB K}}

\def\R{{\rangle}}

\def\d{{\partial}}

\def\be{\begin{equation}}
\def\ee{\end{equation}}
\def\su{\widehat{su}}

\nc{\bra}[1]{\langle {#1}|}
\nc{\ket}[1]{|{#1}\rangle}
\nc{\nn}{\nonumber \\ }
\nc{\al}{\alpha}
\nc{\g}{\gamma}
\nc{\G}{\Gamma}
\nc{\D}{\Delta}
\nc{\eps}{\epsilon}
\nc{\La}{\Lambda}
\nc{\var}{\varphi}
\nc{\pa}{\partial}
\nc{\hf}{\frac{1}{2}}
\nc{\dz}{\frac{dz}{2\pi i}}
\nc{\bin}[2]{\left (\begin{array}{c} {#1}\\ {#2} \end{array}\right )}
\nc{\lra}{\ \leftrightarrow\ }
\nc{\ben}{\begin{equation}}
\nc{\een}{\end{equation}}
\nc{\bea}{\begin{eqnarray}}
\nc{\eea}{\end{eqnarray}}
\nc{\bib}{\bibitem}
\nc{\VM}{M}  
\nc{\RM}{R}  
\nc{\IM}{{\cal I}} 
\nc{\VMR}{\VM^\star}  
\nc{\RMR}{\RM^\star} 
\nc{\ARM}{\RM^{\rm a}} 

\documentstyle[epsfig]{article}
\begin{document}
\title{Logarithmic lift of the $\su(2)_{-1/2}$ model}
\author{F. Lesage$^{a,}$\footnote{lesage@crm.umontreal.ca}\ \ ,\
     P. Mathieu$^{b,}$\footnote{pmathieu@phy.ulaval.ca}\ \ ,\
     J. Rasmussen$^{a,}$\footnote{rasmusse@crm.umontreal.ca}\ \ ,\
     H. Saleur$^{c,d,}$\footnote{saleur@usc.edu}}
\date{\today}
\maketitle

\centerline{${}^a$Centre de Recherches Math\'ematiques,
        Universit\'e de Montr\'eal,}
\centerline{C.P. 6128, succursale centre-ville, Montr\'eal,
      Qc, Canada H3C 3J7}
\vspace{.2cm}
\centerline{${}^b$D\'epartement de Physique, Universit\'e Laval,}
\centerline{Qu\'ebec, Qc, Canada G1K 7P4}
\vspace{.2cm}
\centerline{${}^c$Service de Physique Th\'eorique, CEN Saclay}
\centerline{Gif sur Yvette 91191, France}
\vspace{.2cm}
\centerline{${}^d$Department of Physics, University of Southern California,}
\centerline{Los Angeles, CA 90089-0484, USA}

\begin{abstract}
This paper carries on the investigation of the   non-unitary
$\su(2)_{-1/2}$ WZW model. An essential tool
in our first work on this topic was a free-field representation,
based on a $c=-2$
$\eta\xi$ ghost system, and a Lorentzian boson.
It turns out that there are several `versions' of the $\eta\xi$ system,
allowing different $\su(2)_{-1/2}$ theories. This is explored here in
details.
In more technical terms, we consider extensions (in the $c=-2$ language)
from the small to the large algebra representation and, in a further
step, to the full symplectic fermion theory. In each case, the results are
expressed in terms of $\su(2)_{-1/2}$ representations. At the first new
layer (large algebra), continuous representations appear which are
interpreted in terms of relaxed modules.
At the second step (symplectic formulation),
we recover a logarithmic theory with its characteristic signature, the
occurrence of indecomposable representations.
To determine whether any of these three versions of the
$\su(2)_{-1/2}$ WZW is well-defined, one conventionally
requires the construction of a modular invariant.
This issue, however, is plagued with various difficulties, as we discuss.

\end{abstract}

\section{Introduction}

In a previous work \cite{LMRS}, we studied the
$\beta\gamma$  system and its relations with the
     $\su(2)_{-1/2}$ WZW model.

The $\beta\gamma$ system is a simple and crucial ingredient
in the description of phase transitions in disordered electronic metals,
2D gravity, and string theory. While a `free' theory, a closer look at
its physical properties reveals subtleties. In particular,
we showed in \cite{LMRS} that in the $\su(2)_{-1/2}$ incarnation,
the spectrum of conformal weights is not bounded from below, and operators
with increasingly negative dimensions appear.
This complexity is, however, smoothed out by the fact that all the
operators can be organized in terms of four families,
labeled by the four admissible representations of the  $\su(2)_{-1/2}$
algebra \cite{KW}.

The infinite number of representations within a family are related to each
other through the spectral-flow symmetry. Viewed from the perspective of
the $\beta\gamma$ system, these flowed representations are associated
to what we called deeper twists \cite{LMRS}.

A substantial portion of our previous work was devoted to putting the
equivalence between the  $\beta\gamma$ system and the  $\su(2)_{-1/2}$
WZW model on a firm basis. In particular, we compared the correlators
of the ghost theory with the solutions to the Knizhnik-Zamolodchikov
equation. The faithful free-field representation of the $\beta\gamma$
system, in terms of the  $\eta\xi$ ghosts (with $c=-2$)
and a Lorentzian boson,
played a  crucial role in our analysis of the correlators.

The free-field representation also  allowed us to make
definite statements concerning the fusion rules of the model.
In that regard, comparing our conclusions with those
of another recent work devoted to the analysis of the
$\su(2)_{-4/3}$ WZW model \cite{Gab}, the results may appear puzzling. It
was found there that continuous and indecomposable
(or logarithmic) representations necessarily appear in fusion rules.
In contrast,  we found that the $k=-1/2$ fusion rules close
within the set of usual highest-weight representations and
their flowed versions (which include the corresponding lowest-weight
representations).
In other words, neither continuous nor indecomposable representations are
enforced by closure of the $\su(2)_{-1/2}$ fusion algebra.

The bottom line of this discrepancy is that the $\su(2)_{-4/3}$ WZW model
is believed to be a logarithmic conformal field theory \cite{Gab}, while
the $\su(2)_{-1/2}$ WZW model in its $\beta\gamma$ formulation
appears to be a quasi-rational conformal field theory \cite{LMRS}.
Since the level is admissible in both cases and since admissible
$\su(2)$ models can be expected to
behave similarly, such a different behavior is surprising.

Several resolutions seem to suggest themselves, though.
First, an immediate and quite natural interpretation of the discrepancy
is that the theory could have various formulations, that is, various
{\it lifts}, and that we are actually not comparing the same version
of the two models.

Let us clarify this idea of `various formulations' within the precise
context of the $k=-1/2$ model. The
free-field representation used in \cite{LMRS} is (see also section 3
for a more detailed discussion of this representation):
\ben
        \beta = e^{-i\phi} \eta, \qquad  \gamma = e^{i\phi}\partial \xi\; ,
\een
where $\eta$ and $\xi$ are the $c=-2$ fermionic ghosts of weight
$h=1$ and $h=0$,
respectively. The point we want to stress is that the free-field
representation does not
take advantage of the full $\eta\xi$ algebra; it only uses the
algebra spanned by
$\eta$ and $\pa\xi$, the so-called {\it small algebra} \cite{FMS}.
In other words, we have not used the zero mode of $\xi$.
It is thus natural to investigate the effect, on the spectrum, of
introducing this zero mode. Adding a zero mode to the theory is what
we refer to loosely as a {\em lift}.  Here it is a lift from the
small to the {\it large algebra}.
Although such a lift does not affect the central charge, it certainly
generates additional states.

Actually, the extra zero mode is responsible for the occurrence of new
{\em continuous} representations. An analogous example is the
representation denoted $E$ in \cite{Gab}.
Here we show that, in the more general case, they correspond collectively to
the so-called {\it relaxed  representations} of \cite{FST} (whose
structure is briefly reviewed in the appendix).

But this lifting process may be pushed a step further. If we
extend the dimension-one ghost
$\eta$ by a zero mode, more precisely the zero mode of $\pa^{-1}\eta$,
one would then have a further lift. This corresponds to a
representation in terms of the $c=-2$ symplectic fermion theory or {\it
symplectic algebra}. At this stage, there are thus two zero modes
and the corresponding $c=-2$ theory is known to be
logarithmic. This characteristic  will obviously be shared by the parent
$c=-1$ model.  The symplectic formulation thus leads to a candidate
logarithmic version of the $\su(2)_{-1/2}$ WZW model.

In fact, by extending the theory by two fermionic zero-modes, it is clear that
every bosonic state can thereby be paired with a bosonic partner. The
latter is obtained simply by  acting  upon
the original state with the product of these two fermionic zero modes.
This is a candidate for a two-dimensional Jordan cell -- an indicator
of a logarithmic theory.  This contention is indeed
confirmed and we find  that the resulting theory is now very similar
to that analyzed in \cite{Gab}. In particular, we see the indecomposable
representations emerging and the rather complicated pattern of
their `extremal diagram' is made very concrete by our free-field
computations. In this context, we make the  new observation that the
indecomposable representations have two constituent relaxed modules.

A natural question at this stage is the following: are these various
versions of the $\su(2)_{-1/2}$ WZW model all well-defined theories
or, say, is the logarithmic version  the only
viable one (as the analysis of \cite{Gab} would suggest)? In all
versions, we show that there is a closed fusion algebra.
However, one may be dealing merely with a closed {\it subalgebra} at each
step of some `larger' theory. For the Ising model, for instance, the
fusion rules close within the Neveu-Schwarz sector. However, the
physical model has to include the Ramond sector as well. In such
a circumstance, the presence of extra fields becomes apparent at
the level of the construction of the modular invariant.
Therefore, these questions point toward the analysis of the modular
invariant partition function. Is a modular invariant meaningful for
any of these versions (i.e., can we construct three different modular
invariants for the $\su(2)_{-1/2}$ WZW model) or is there a
single `master' modular invariant associated to the logarithm
version of the theory?

Somewhat surprisingly, the modular invariant issue is plagued with
various difficulties. Although we do not propose definite answers, some of the
subtleties related to this problem are pointed out in section 5.

The paper is organized as follows.
First we discuss purely algebraic structures
of the $\su(2)_{-1/2}$ algebra without reference to the free fields.
We discuss the spectral flow symmetry and the different standard modules.
The extension to relaxed modules and
some of their embedding patterns is presented in the appendix, which can
be read after section 2. We then analyze in turn the extension of the
theory from the small to the large algebra (section 3) and then its
extension to the symplectic  formulation (section 4). The problems linked
to the interpretation of the modular invariant are addressed in section 5.

\vskip0.3cm
\noindent {\bf Notation}:

\noindent For an easy reference, we summarize our notation for the
various types of
modules used here (which will be defined in due time):\\[.1cm]
$\VM$: highest-weight Verma module;\\
       $\VMR$: irreducible highest-weight module;\\
$\RM$: relaxed module;\\
       $\RMR$: irreducible relaxed module;\\
$\ARM$: almost reduced relaxed module;\\
$\IM$: indecomposable module.

\section{Admissible modules, flows and fusion}

In this section, we review results concerning the spectral
flow and the related twisted modules or representations.
Some of these results were already reported in \cite{LMRS} but here the
emphasis is placed somewhat differently. Moreover, we will attempt
to place our results within a more general algebraic framework
discussed in particular in \cite{FST,SS}. For this,
it will be convenient to slightly modify our previous notation and
adjust it to the ones of these references.
After this short review, we relate
the spectral flow to a symmetry of the fusion rules. We then  provide a simple
characterization of those models for which all fusion rules can be
obtained by flows of the fusion of level-zero  finite-dimensional
representations. This provides a simple distinction between the
level $-1/2$ and the level $-4/3$ cases.

\subsection{Spectral flow and twisted modules}

The affine Lie algebra $\su(2)_k$ with level $k$
is defined by the commutator relations
\bea
        \left[J^+_m,J^-_n\right]&=&2J^3_{m+n}+km\delta_{m+n,0}\ ,\nn
        \left[J^3_m,J^\pm_n\right]&=&\pm J^\pm_{m+n}\ ,\nn
        \left[J^3_m,J^3_n\right]&=&\frac{k}{2}m\delta_{m+n,0}\; .
\label{su2}
\eea
The spectral flow \cite{BH,FST,SS},
with flow parameter\footnote{In our previous work \cite{LMRS} we used the
inverse flow, acting as $\pi_w:\ J^+_n\mapsto J^+_{n-w}$, for example.}
$\theta\in\ZZ$, is the algebra automorphism
\ben
        \pi_\theta\ :\ \ \ \ \
         J^\pm_n\ \mapsto\ J^\pm_{n\pm\theta},\ \ \ \ J^3_n\ \mapsto\
         J^3_n+\frac{k}{2}\theta\delta_{n,0}\; .
\label{flow}
\een

A twisted Verma module, $\VM_{j,t;\theta}$, is freely generated
by $J^+_{n\leq\theta-1}$, $J^3_{n\leq-1}$, and $J^-_{n\leq-\theta}$
from the twisted highest-weight vector $\ket{j,t;\theta}$ defined by the
conditions
\bea
        J^+_{n\geq\theta}\ket{j,t;\theta}\ \ =\ \ 
J^3_{n\geq1}\ket{j,t;\theta}&=&
         J^-_{n\geq1-\theta}\ket{j,t;\theta}\ \ =\ \ 0\; ,\nn
        \left(J^3_0+\frac{k}{2}\theta\right)\ket{j,t;\theta}
&=&j\ket{j,t;\theta} \; .
\label{verma}
\eea
$j$ is referred to as the spin of the vector $\ket{j,t;\theta}$.
In order to comply with the notation of \cite{SS} we use $t\equiv k+2$
in the characterization of the module and its vectors. We assume $t\neq0$.
When the twist parameter $\theta\in\ZZ$ is omitted, it is understood to be
zero and the module or vector is not twisted.

A vector in the twisted module $\VM_{j,t;\theta}$ can be assigned the
level $N$ if it can be written as a linear combination of monomials
\ben
        J_{-n_1}^{a_1}\dots J_{-n_i}^{a_i}\ket{j,t;\theta}\; ,
\een
having $\sum_{j=1}^in_j=N$. Note that some $n_j$ may be negative.
Such a vector has $J^3_0$ eigenvalue of the form
\ben
        m\in \left(j-\frac{k}{2}\theta+\ZZ\right)\; .
\label{m}
\een
This eigenvalue is called the charge of the vector.
Within a module,  we choose to talk about the level of a vector
instead of its conformal
weight, i.e.,  its $L_0$ eigenvalue where $L_0$ is the zero mode of
the Sugawara energy-momentum tensor $T$. The rationale for this is
that the notion
of level is still well-defined for indecomposable representations, to be
discussed below, while the action of $L_0$ generally is non-diagonal.

The vectors in $\VM_{j,t;\theta}$ (as well as in the more complicated
indecomposable modules) can always be
organized in linear combinations having definite charge and level.
The representation of $\VM_{j,t;\theta}$ by a semi-infinite
square lattice in the $(charge,-level)$ plane is called a diagram
(we used similar diagrams, albeit with different conventions,
in Fig. 1 of our previous paper \cite{LMRS}). As indicated, the second axis
is commonly inverted. A vertex represents the finitely many vectors
with given $(m,-N)$ values. From (\ref{m}) and the definition of level,
it follows that the lattice is integer spaced, and shifted from the
integer points by the vector $(j-\frac{k}{2}\theta,0)$. It is natural
to make an additional shift by $(0,h_{j,\theta})$,
where $h_{j,\theta}$ is the conformal dimension of the twisted
highest-weight state $\ket{j,t;\theta}$:
\ben
      h_{j,\theta}= {j(j+1)\over k+2}-j\theta+ {k\over 4}\theta^2\;.
\een
Vertices are connected by arrows if the corresponding
vectors can be reached by the action of the
$\su(2)_k$ generators. The extremal diagram is the boundary of the lattice
and is often the only part illustrated.

\begin{figure}
\centerline{\epsfig{figure=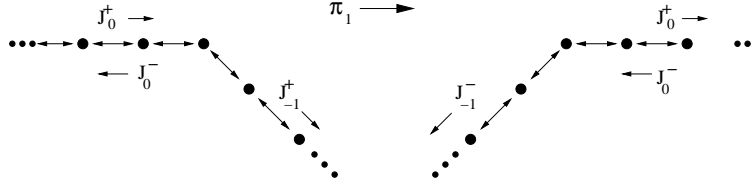,width=10cm}}
\vspace{0.3cm}
\caption{Diagram representations of $M_{j,t;\theta}$, where the horizontal
axis is the charge, the vertical axis the level. The `extremal diagram'
on the left represents an untwisted highest-weight module.
The corresponding (highest-weight) $su(2)$ representation at level zero
may extend infinitely towards the left. $\pi_1$ maps this module onto
a lowest-weight module, represented by the diagram on the right.}
\label{Fig.pi}
\end{figure}

With its action extended to Verma modules,  $\pi$ induces the following
translation
\ben
        \pi_\theta(\VM_{j,t;\theta'})=\VM_{j,t;\theta+\theta'}\;.
\label{trans}
\een
The flow and twist parameters are then identified.
It follows from (\ref{trans}) that (twisted) Verma modules may be
organized in orbits under the spectral flow. Under the action of
$\pi_1$, a highest-weight module $\VM_{j,t}$ is sent to a lowest-weight
(that is, lowest weight with respect to the finite $su(2)$ algebra
acting at level zero) module $\VM_{j,t;1}$, see Fig. \ref{Fig.pi}.
Thus, in every orbit there are at
least one highest-weight and one lowest-weight module. As we will
discuss shortly, some modules are both highest-weight and lowest-weight
modules.

\subsection{Reducible modules and admissible representations}

A singular vector exists in $\VM_{j,t}$ if and only if the spin $j$ takes
       one of the two forms
\bea
        &&j=j^+(r,s,t)=\frac{r-1}{2}-\frac{s-1}{2}t,\ \ \ \ r,s\in\NN\;,\nn
        &&j=j^-(r,s,t)=-\frac{r+1}{2}+\frac{s}{2}t,\ \ \ \ \ \ \ r,s\in\NN\;.
\label{j+-}
\eea
The singular vectors in $\VM_{j^\pm(r,s,t),t}$ have been worked out
explicitly by Malikov, Feigin and Fuchs \cite{MFF}
and are referred to as MFF vectors. They appear at levels $r(s-1)$
and $rs$, respectively.

Due to the presence of singular vectors, some modules are reducible.
Their irreducible counterparts are obtained by factoring out the
submodules generated from the singular vectors.
At most two of the MFF vectors are primitive, meaning that we need
to factor out vectors generated from at most two singular vectors.
The irreducible module associated to the Verma module $\VM_{j,t}$
will be denoted $\VMR_{j,t}$.

For fractional or integer $t=k+2>0$
\ben
        t=p/p',\ \ \ \ \ \ (p,p')=1\;,
\een
the admissible spins are of the form $j^+$ appearing in (\ref{j+-})
and subject to the constraints
\ben
        j(r,s)=\frac{r-1}{2}-\frac{s-1}{2}t,\ \ \ \
         1\leq r\leq p-1,\ \ 1\leq s\leq p'\; .
\label{adm}
\een
For the case $t=3/2$ ($k=-1/2$) of our previous paper \cite{LMRS},
we have $p=3$ and $p'=2$. This corresponds to the $\beta\gamma$ system,
and there are thus four admissible spins:
\ben
     j(1,1)=0,\qquad j(2,1)=1/2,\qquad j(2,2)=-1/4,\qquad j(1,2)=-3/4\;.
\een
For generic $t$, the total number of admissible spins is $(p-1)p'$.

A representation is called admissible if it corresponds to a
highest-weight module, $\VM_{j,t}$, with $j$ an admissible spin.
The horizontal extremal diagram, i.e.,  the $su(2)$ module generated by the
action of the zero modes on $\ket{j,t}$, is finite-dimensional
if and only if $s=1$ (it is otherwise semi-infinite). An irreducible Verma
module $\VMR_{j(r,1),t}$ may be regarded as a highest-weight and a
lowest-weight module simultaneously. We sometimes loosely  call the $p-1$
modules $\VMR_{j(r,1),t}$ finite-dimensional, even though this
designation applies to their horizontal extremal diagrams only.
We call an orbit containing an admissible highest-weight module an
admissible orbit.

As already mentioned, $\pi_1(\VMR_{j(r,s),t})$ is a lowest-weight module.
In order for it to be a highest-weight module as well,
the vector $(J^+_{-1})^u\ket{j(r,s),t}$ in the extremal diagram of
$\VM_{j(r,s),t}$ must be singular (see Fig. \ref{Fig.pi}).
This is seen to require
\ben
        u=k-2j+1=-r+st\;,
\een
and since $u$ is a positive integer, we find that $s=p'$.
In terms of reduced modules, we conclude that the only orbits with
more than one highest-weight module are the ones containing
the finite-dimensional modules. For fractional
level\footnote{We are not concerned with the simpler case of integer
levels (where $p'=1$).}, $p'>1$, these orbits contain exactly two
highest-weight modules, and the latter are related as
\ben
        \pi_{-1}(\VMR_{j(r,1),t})=\VMR_{j(r,p'),t} \;.
\een
In particular, two finite-dimensional modules cannot belong to the
same orbit. A simple counting shows that there are $(p-1)(p'-1)$
admissible orbits (in the case with $p=3$ and $p'=2$ there are thus
two admissible orbits).
We note that the number of finite-dimensional modules is equal to
the number of admissible orbits if and only if $p'=2$. This is
of importance for the closure of the fusion algebra to be discussed
in the following subsection.

\subsection{Closure of the fusion algebra}

For simplicity, when discussing fusions we shall label the fused
fields by their associated modules, e.g., we may consider the fusion
$\VMR_{j,t;\theta}\times \VMR_{j',t}$.
Following \cite{Gab,LMRS}, we assume that fusion products are invariant
under the spectral flow:
\ben
      \pi_\theta(\VMR_1)\times\pi_{\theta'}(\VMR_2)
       =\pi_{\theta+\theta'}(\VMR_1\times \VMR_2)
\label{pifusion}
\een
for some irreducible modules $\VMR_i$. Later on we shall extend
this assumption to include relaxed and indecomposable modules as well.
It follows from (\ref{pifusion})
that the fusion product of two modules is determined
by the fusion product of the highest-weight modules in their
respective orbits.

Let us now consider the case where fusion closes on the set of
finite-dimensional modules or representations, i.e., on the set of
admissible irreducible modules $\VMR_{j(r,1),t}$.
In the integer-level
case, this is certainly always the case, while in the fractional-level
case it remains
an assumption in general. However, for $\su(2)_{-1/2}$, this has been
demonstrated explicitly  using a free-field realization \cite{LMRS}.

Now, whenever fusion closes on the set of
finite-dimensional modules (and as we just said, this is so for
$k=-1/2$) and for all cases where $p'=2$ (which includes $k=-1/2$), it
follows that the complete fusion algebra is fixed by the fusion algebra of
the finite-dimensional representations. Closure of the fusion algebra is
thereby ensured (as for $k=-1/2$, for example).
On the other hand, we also see that the fusion algebra is
not fully determined by the fusion algebra of the finite-dimensional
representations when $p'>2$. It may still close, i.e., close on the
set of $\VMR_{j,t;\theta}$, but the analysis of the $k=-4/3$ example
in \cite{Gab} suggests the contrary. We find it thus natural to
conjecture that the fusion algebra closes on the set of admissible
representations and their flowed companions if and only if $p'=2$.

For $p'=2$, it turns out that the fusion algebra of twisted modules,
$\VMR_{j(r,s),p/2;\theta}$, closes on the subset defined by
$\theta\in\ZZ_\geq$ (or the subset defined by $\theta\in\ZZ_\leq$) as well.
Of course, closure of the fusion algebra is only a necessary
requirement for a well-defined theory; among other requirements,
modular invariance of the partition function must be addressed. We
get back to  this point in section 5.

\subsection{Characters of representations as series expansions of
character functions}

Let us now describe the irreducible  characters of the admissible
representations and the associated concept of character functions.

To each twisted admissible representation there is an associated character,
$\chi_{j(r,s),t;\theta}$, defined by
\ben
        \chi_{j(r,s),t;\theta}(z,q)={\rm Tr}_{\VMR_{j(r,s),t;\theta}}
         q^{L_0-c/24}z^{J^3_0} \;,
\label{char}
\een
where $z$ and $q$ are formal parameters. In the following we will
consider functions of $z$ and $q$ and
assume that $|q|<1$, allowing us to consider the constraint
as a $q$-dependent region in the $z$ plane.
In order to comply with the notation of \cite{FSST},
we have changed our notation from $y$ in \cite{LMRS}
to $z=y^2$. We stress that the characters (\ref{char}) are formal
series in the parameters $q$ and $z$. As discussed in \cite{FSST},
these series only converge in certain domains (annuli)
of the complex plane, while they can be continued outside. The resulting
meromorphic functions are called character functions, and must not be
confused with the original characters. We will denote these character
functions $F_{j(r,s),t;\theta}(z,q)$.\footnote{To simplify the notation,
we will often use the notation
$F_{j(r,s)}(z,q)=F_{j(r,s),\frac{3}{2};0}(z,q)$ for the four distinct
character functions in the case $k=-1/2$.}

We stress that $F_{j(r,s),t;\theta}(z,q)$ is holomorphic if and only if
$s=p'=1$, i.e., for integrable representations.

The functions $F_{j(r,s),t;\theta}(z,q)$ may be organized in finitely
many  orbits. To see this, we trivially extend the definition of the
spectral flow to the character functions as
\ben
        \pi_{\theta'}(F_{j(r,s),t;\theta})(z,q) =
F_{j(r,s),t;\theta+\theta'}(z,q) \;,
\een
where
\ben
     F_{j,t;\theta}(z,q)=q^{\frac{t-2}{4}\theta^2}z^{-\frac{t-2}{2}\theta}
      F_{j,t;0}(zq^{-\theta},q)\;.
\een
Using the relations (cf. \cite{FSST}, Theorem 4.2)
\bea
     \pi_{2p'}(F_{j(r,s),t;\theta})(z,q)&=&F_{j(r,s),t;\theta}(z,q) \;,\nn
     \pi_{p'}(F_{j(r,s),t;\theta})(z,q)&=&(-1)^{p'-1}
      F_{j(p-r,s),t;\theta}(z,q)\;,
\label{rel}
\eea
one observes that only a finite number of functions are generated
under the spectral flow.\footnote{There are additional relations when
$p=2r$ (see \cite{FSST} for details).} In the case with $k=-1/2$,
where $t=p/p'=3/2$, the relations (\ref{rel}) lead to a periodicity
four in the set of character functions. This periodicity has also
been noticed in section 6.3 of \cite{LMRS}.

\begin{figure}
\centerline{\epsfig{figure=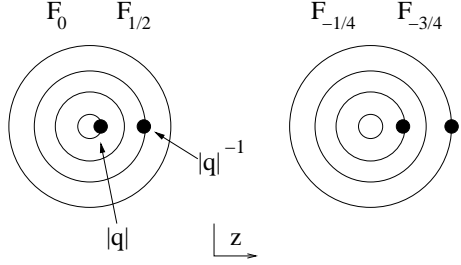,width=6cm}}
\vspace{0.3cm}
\caption{Schematic description of the distribution of poles.}
\label{Fig.poles}
\end{figure}

The character functions are meromorphic in $z$; their singularities
are all simple poles. The
distribution of these poles is as follows: for $j=j(r,s)=j^+(r,s,t)$,
the function $F_{j}(z,q)$ has poles
at the points (cf. \cite{FSST}, Lemma 4.4)
\ben
        z=q^n,\ \ \ \ \ \ \ n\in\ZZ\setminus(p'\ZZ+s-1)\; .
\label{poles}
\een
For $p'=2$, this leads to a pole on every second circle
defined by $|z|=|q^n|$. The pole structure of the four character
functions appearing for $k=-1/2$ is illustrated in Fig. \ref{Fig.poles}.
Let us describe it in more details. For $s=1$ we have a pole located
at every odd power of $q$. Consider for instance the
character of the untwisted ($\theta=0$) identity representation with
$j(1,1)=0$. The series expansion was originally defined in the
annulus $ 1<|z|<|q|^{-1}$. However, since there is no
pole on the unit circle, this expansion is extended to an open
`double  annulus' $ |q| < |z| < |q|^{-1}$.\footnote{We stress that
the extension of the convergence domain to two basic annuli
is a special feature of the $p'=2$ models. The situation  for all
fractional levels with $p'=2$ is easily described:
poles appear exactly after every second annulus.
The generalization of Fig. \ref{Fig.poles} is straightforward:
the $p-1$ character functions associated to finite-dimensional
representations, $j(r,1)$, have identical distributions (extending
the left part of the figure), while the $p-1$ infinite-dimensional
representations, $j(r,p')$, give rise to the functions
with distribution as to the right in Fig. \ref{Fig.poles}.}
The character of the untwisted identity representation thus reads
\ben
      \chi_{j(1,1),\frac{3}{2};0}(z,q)= F_{0}(z,q)\qquad
{\rm iff} \ \ |q| < |z| < |q|^{-1}\;,
\een
with
\begin{equation}
       F_{0}=q^{1/24}{\sum_{m\in Z}q^{6m^2-2m}z^{-3m}-z^{-1}\sum_{m\in
        Z}q^{6m^2+2m}z^{-3m}\over
        \prod_{m\geq 0}\left(1-z^{-1}q^{m}\right)
        \prod_{m\geq 1}\left(1-zq^{m}\right)\left(1-q^{m}\right)}\;.
        \end{equation}
A similar result holds for the spin-$1/2$ representation with
\ben
       F_{1/2}=q^{1/24}q^{1/2}z^{1/2}{\sum_{m\in Z}
       q^{6m^2-4m}z^{-3m}-z^{-2}\sum_{m\in
        Z}q^{6m^2+4m}z^{-3m}\over
        \prod_{m\geq 0}\left(1-z^{-1}q^{m}\right)
        \prod_{m\geq 1}\left(1-zq^{m}\right)\left(1-q^{m}\right)}\;.
\een
In the same way, the character of the $j(2,2)=-1/4$
representation is found to be
\ben
        \chi_{j(2,2),\frac{3}{2};0}(z,q)=F_{-1/4}(z,q) \qquad
          \ \ {\rm iff} \ \ 1 < |z| < |q|^{-2}\;,
\een
with
     \begin{equation}
         F_{-1/4}=q^{1/24}q^{-1/8}z^{-1/4}{\sum_{m\in
         Z}q^{6m^2-m}z^{-3m}-z^{-2}q^{2}\sum_{m\in
         Z}q^{6m^2+7m}z^{-3m}\over
         \prod_{m\geq 0}\left(1-z^{-1}q^{m}\right)
         \prod_{m\geq 1}\left(1-zq^{m}\right)\left(1-q^{m}\right)}\;,
         \end{equation}
and similarly for the spin $j=-3/4$ with
\begin{equation}
        F_{-3/4}=q^{1/24}q^{-1/8}z^{-3/4}{\sum_{m\in
        Z}q^{6m^2+m}z^{-3m}-z^{-1}q\sum_{m\in
        Z}q^{6m^2+5m}z^{-3m}\over
        \prod_{m\geq 0}\left(1-z^{-1}q^{m}\right)
        \prod_{m\geq 1}\left(1-zq^{m}\right)\left(1-q^{m}\right)}\;.
        \end{equation}

Consistency with the spectral flow follows from simple identities
between the character functions. One such identity is
\ben
        \pi_{-1}(F_{j(1,1),\frac{3}{2};0})(z,q)
         =F_{j(1,1),\frac{3}{2};-1}(z,q) = q^{-\frac{1}{8}}
z^{-\frac{1}{4}} F_{0}
         (zq,q) = F_{-1/4}(z,q)\;.
\een
Note that both functions can be expanded to the desired characters only
in the annulus $1<|z|<|q|^{-1}$.

The presence of poles in highest-weight representations with $s\not=1$ is
rooted in that at level zero there is an infinite
number of states simply because the representation is
not also lowest weight. That explains, for instance, the fact that
$F_{-1/4}$ has a pole at
$z=1$. But how can we see the poles in $F_0$, say?  For the vacuum
representation there is no pole at $z=1$ because the number of states at
each level is finite, as for an integrable representation. However, it
differs from an integrable representation in being `wider' at each level.
In particular, we have $(J^-_{-1})^n I\not=0$ for any $n$ (i.e., on the
$45^o$ N-W diagonal there are states at every point -- in sharp contrast
to an integrable representation). This implies that if we fix $z=q$,
whose effect is that all states on this diagonal are sent to level
zero, the character becomes infinite.

The divergence of the characters corresponds to a divergence of the
functional integral defining the $\beta\gamma$ system (see our first
paper \cite{LMRS}) when a field is coupled to the $U(1)$ charge.
Analytic continuation beyond the circles of convergence can be seen as
a regularization of this integral.
This will be discussed more in  section 5.

\section{The $\su(2)_{-1/2}$ model revisited}

In the following sections we concentrate on the $\su(2)_{-1/2}$ model.
We discuss free-field realizations of the small and large $c=-2$ algebras.
To keep the notation simple, we leave out the dependency on $t=k+2=3/2$,
as in $\VM_{j;\theta}=\VM_{j,3/2;\theta}$, for example.

\subsection{Free-field realization of the small algebra}

In our previous paper \cite{LMRS} we discussed the  $\beta\gamma$ system,
described by the energy-momentum tensor\footnote{Here and in the following
we omit normal-ordering signs whenever confusion is unlikely to occur.}
\ben
        T=\frac{1}{2}(\beta \partial\gamma-\partial\beta\gamma)
\een
in the left-moving sector. In this formulation, $\beta$ and
$\gamma$ have weight $h=1/2$ and charge $1/2$ and $-1/2$,
respectively, with respect to the current
\ben
        J^3 = -\frac{1}{2} \gamma\beta\;.
\een
The full $\su(2)$ symmetry emerges through the construction
of the currents as
\bea
        J^+ &=& \frac{1}{2} \beta^2 \;,\nn
        J^3&=& -\frac{1}{2} \gamma\beta\;,\nn
        J^-&=& -\frac{1}{2} \gamma^2\;.
\eea
A more convenient representation of these currents is obtained
by introducing
\ben
        \beta = e^{-i\phi} \eta, \ \  \gamma = e^{i\phi}\partial \xi\;,
\een
where $\phi$ is a free boson with negative metric
\ben
        \langle\phi(z)\phi(w)\rangle = \ln(z-w)\;,
\een
while $\eta$ and $\xi$ are fermions of weight $h=1$ and $h=0$,
respectively, satisfying
\ben
        \langle\eta(z)\xi(w)\rangle = \frac{1}{z-w}\;.
\een
In this representation, we have\footnote{The minus sign in $J^-$
corrects a misprint in
\cite{LMRS}.}
\bea
        J^+ &=& \frac{1}{2} e^{-2i\phi} \pa\eta \eta \;,\nn
        J^3 &=& \frac{i}{2}\pa\phi\;,\nn
        J^- &=& -\frac{1}{2} e^{2i\phi} \pa^2\xi \pa\xi\;.
\eea
This free-field representation  provides a
faithful description of the relevant modules in the theory.
For example, the irreducible modules
$\VMR_{0;\theta}$ and $\VMR_{1/2;\theta}$ for small values of $|\theta|$
are represented by the components \cite{LMRS}:
\bea
        \VMR_{0;0} &:& \  \{ 1 \} \;,\nn
        \VMR_{0;1} &:& \ \{ e^{-i\phi/2} \} \;,\nn
        \VMR_{0;-1}\ =\ \VMR_{-1/4;0} &:& \ \{ e^{i\phi/2} \} \;,\nn
        \VMR_{1/2;0} &:& \ \{ e^{-i\phi} \eta \} \;,\nn
        \VMR_{1/2;1} &:& \ \{ e^{-3i\phi/2} \eta \} \;,\nn
        \VMR_{1/2;-1}\ =\ \VMR_{-3/4;0} &:& \ \{ e^{3i\phi/2} \partial\xi
\}\;.
\label{repM}
\eea
Any vector in the irreducible module is generated from the above
representative by the action of the $J^a_n$ modes as given by their free-field
expressions.
The resulting modules may be identified with well-known representations.
In the same order as listed above, and in the notation of \cite{LMRS},
we have $D_0$, $D^-_{-1/4}$, $D^+_{-1/4}$, $D_{1/2}$, $D^-_{-3/4}$, and
$D^+_{-3/4}$.

In this free-field picture, the action of the spectral flow amounts to
a multiplication by a vertex operator:
\ben
        \pi_\theta(\Psi) \ \rightarrow \  :\Psi e^{-i\theta \phi/2}: \;.
\label{vertex}
\een
It is simple to verify that the fusion algebra of all these fields closes.

The currents $J^a$ involve the field $\xi$ only through
its derivative, and this  is also true for the representations described
above, (\ref{repM}): in the terminology of \cite{FMS,Kau}, we are
working with the small algebra.

\subsection{Extension to the large algebra}

Extending the theory by including $\xi$ amounts to studying the large
algebra instead of the small algebra discussed above. By this extension,
the possibility of new and possibly reducible modules emerges.  In order
to see that such modules actually appear, let us study the representative
\ben
        \xi e^{i\phi/2} \;.
\label{vec1}
\een
By acting with $J^+$ we find
\bea \label{actdejplus}
        J^+(z)\ (e^{i\phi/2}\xi)(w)&=&\hf e^{-2i\phi} \pa
         \eta\eta(z) \ e^{i\phi/2}\xi(w) \nn
        &=&\frac{1}{(z-w)}\left( \frac{1}{2} e^{-3i\phi/2}
         \eta\right) (w) + \cdots\;,
\eea
and similarly
\bea
        J^-(z)\ (e^{i\phi/2}\xi)(w)&=&-\hf e^{2i\phi} \pa^2
         \xi\pa\xi(z) \ e^{i\phi/2}\xi(w) \nn
        &=&\frac{1}{(z-w)}\left( -\frac{1}{2} e^{5i\phi/2}
         \pa^2\xi\pa\xi\xi\right) (w)+ \cdots\;.
\eea
Here and in the following we shall make frequent use of the expansions
\bea
        J^a(z)&=&\sum_{n\in\ZZ}\frac{J_n^a(w)}{(z-w)^{n+1}} \;,\nn
        T(z)&=&\sum_{n\in\ZZ}\frac{L_n(w)}{(z-w)^{n+2}}\;,
\eea
around the point $w$ of the affine currents and the energy-momentum tensor.
In particular, we see from (\ref{actdejplus}) that
\bea
     J_0^+ (e^{i\phi/2}\xi)=\frac{1}{2} e^{-3i\phi/2} \eta\;.
\eea
On the other hand, this last expression
corresponds to the lowest-weight state with $J^3_0$ eigenvalue $3/4$ of the
lowest-weight representation $\VM_{1/2;1}^*$. The extremal diagram of the
full representation, i.e.,  the one obtained by acting with the current
generators on $\xi e^{i\phi/2}$, is depicted in Fig. \ref{large1}.
(Note that it differs from Fig. \ref{charged} in the appendix.)
\begin{figure}
\centerline{\epsfig{figure=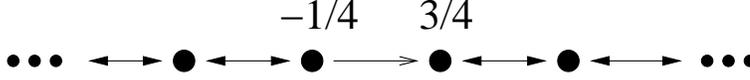,width=10cm}}
\vspace{0.3cm}
\caption{Extremal diagram for the relaxed module generated from
        $\xi e^{i\phi/2}$.}
\label{large1}
\end{figure}

As shown in Fig. \ref{large1}, the representation is reducible.
The Verma module associated with this representation is that of
a relaxed module containing a charged singular vector in the terminology
of \cite{FST,SS}.\footnote{For example, using the
notation introduced in  the appendix, we can label the vector with
eigenvalue
$J_0^3=-1/4$ by $\ket{\mu_1,\mu_2;\theta}=\ket{0,-1/2;0}$.}

Let us pause to explain in simple terms the essence of relaxed and almost
reduced modules (formal definitions can be found in the appendix). An
untwisted
relaxed module, denoted $\RM$, is an affine Lie algebra highest-weight
module except for the fact that the $J^+_0$ highest-weight condition is
no longer imposed.  As a result, if $j$ is not half-integer (meaning that
the level-zero $su(2)$ representation is not lowest-weight, in particular),
the representation at level zero is infinite in both directions. A relaxed
module may contain singular vectors in which case the
corresponding irreducible module (written $\RMR$) is obtained by
factoring out these singular vectors. In a relaxed module, there may be a
(special) singular vector at level zero. It is called a charged
singular vector.
If all but this special singular vector are factored out,
the module is said to be almost reduced and is denoted $\ARM$.
Relaxed modules without charged singular vectors are discussed
in \cite{Gab,LMRS} and are there denoted $E$ and are referred
to as continuous representations (see also the appendix).
As for ordinary modules, relaxed modules may be twisted.

We now compare the large-algebra module just obtained (and generated
from (\ref{vec1})) with those generated by the small algebra. The main
difference  is that
this new module, described by the free fields in the large algebra,
is {\em not} the irreducible module
obtained by removing all singular vectors. Rather, it is a reducible module
obtained by removing all singular vectors but the charged singular one
in the extremal diagram.  The presence of this singular vector is
signaled in Fig. \ref{large1} by the absence of an arrow going from
$m=3/4$ to $m=-1/4$. It is thus an example of an almost reduced
module defined above.

We have discussed explicitly the structure of the module
at the highest (extremal) level.
Let us show that it has a similar pattern at lower levels as well.
We start from the field $\xi e^{i\phi/2}$.
By observing that the operator $e^{i\phi/2}$
is generating an irreducible $j=-1/4$ highest-weight
representation, we know that the extremal diagram is
such that $J_{-1}^+ e^{i\phi/2}=0$,
for example.  Similar conditions appear at lower levels.
On the other hand, acting separately on $\xi$ and $e^{i\phi/2}$,
we observe that for every point where an annihilation
condition is met for $e^{i\phi/2}$, one can contract one of
the $\eta$'s in $J^+$ with
$\xi$ to create a pole (or double pole) nullifying the extremal condition.
Every time this is done, the zero mode of $\xi$ is lost, and a
similar arrow, as that at the top level, is created -- we are now
in a state that is part of the lowest-weight module with $J^3_0$
eigenvalues in $\ZZ+3/4$.

We note that this extension to an almost reduced module
is consistent with the `equations of motion'  since
$C=-3/16$ (following from $|\mu_1-\mu_2|=1/2$, see (\ref{mu1mu2})). From
(\ref{vertex}) it follows straightforwardly that $\xi$ and
$\xi e^{-i\phi/2}$, for example, correspond to states in 
modules\footnote{The notation is $\ARM_{\mu_1,\mu_2;\theta}$
where $\mu_1$ and $\mu_2$ are described in the appendix.
The module $\ARM_{0,-1/2;0}$ is the one in Fig. \ref{large1}.},
$\ARM_{0,-1/2;\theta}$, obtained under the spectral
flow of $\ARM_{0,-1/2;0}$. Their characters are given by
\ben
        \chi^{\ARM}_{0,-1/2;\theta}(z,q) = \delta(zq^{-\theta},1)
       Res_{zq^{-\theta}=1} \left[ q^{-\theta^2/8}
         z^{\theta/4} F_{-1/4}(zq^{-\theta},q) \right] \; ,
\label{0-12}
\een
where the delta function is defined as follows
\ben
\delta(u,1) = \sum_{n\in \ZZ} u^n\;.\label{delu1}
\een
In the untwisted case, it reflects the fact that the spectrum of
states at level
zero is infinite in both directions. One finds
\begin{equation}
     \chi^{\ARM}_{0,-1/2;\theta}(z,q) = \delta(zq^{-\theta},1)
      q^{-1/12}q^{-\theta^2/8}\prod_{n=1}^\infty {1\over (1-q^n)^2}\;,
\end{equation}
where we have used the identity\footnote{This identity can be proven, 
for instance, by some simple manipulations of the Jacobi $\theta_4$ function.}
\ben
        \sum_{m\in Z}q^{6m^{2}-2m}q^{3m(2p+1)}-q^{2p+1}
        \sum_{m\in Z}q^{6m^{2}+2m}q^{3m(2p+1)}=
        q^{-p(3p+1)/2}\prod_{n=1}^{\infty}(1-q^{n}),\ \ \ ~(p\in\ZZ)\;.
        \een

\begin{figure}
\centerline{\epsfig{figure=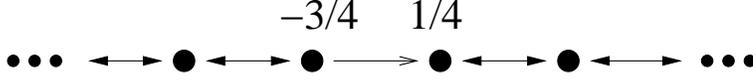,width=10cm}}
\vspace{0.3cm}
\caption{Extremal diagram for the relaxed module generated from
        $e^{3i\phi/2}\xi\partial\xi$.}
\label{large2}
\end{figure}

There is, however, another module which is not in the orbit of
$\ARM_{0,-1/2;0}$. It is generated from the representative
\ben
        e^{3i\phi/2} \partial\xi \xi\;.
\een
With an analysis similar to the previous one, we obtain the extremal
diagram described in Fig. \ref{large2}.
The characters of the almost reduced modules in the orbit of this one
read
\ben
       \chi^{\ARM}_{0,1/2;\theta}(z,q) = \delta(zq^{-\theta},1)
       Res_{zq^{-\theta}=1}\left[ q^{-\theta^2/8}
        z^{\theta/4} F_{-3/4}(zq^{-\theta},q) \right] \;,
\label{012}
\een
and one finds again
\begin{equation}
      \chi^{\ARM}_{0,1/2;\theta}(z,q) = \delta(zq^{-\theta},1)
      q^{-1/12}q^{-\theta^2/8}\prod_{n=1}^\infty{1\over (1-q^n)^2}\;.
\label{ra0}
\end{equation}

In both examples, (\ref{0-12}) and (\ref{012}), we have chosen to characterize
the relaxed module by the labels, $\mu_1$ and $\mu_2$, defining the
neighbour vector to the left of the charged singular vector
(see Fig. \ref{large1} and Fig. \ref{large2}). This characterization
is of course not unique, as every vector to the left of the charged
singular one could play the role as the relaxed highest-weight vector.
It is only in the fully reduced modules that these two choices
are singled out.

The extension from the small to the large algebra is asymmetric
since we chose to include the zero mode of $\xi$.
The asymmetry shows up in the fact that only relaxed extensions
of highest-weight representations are present.
We could as well have introduced the zero mode of `$\pa^{-1}\eta$'.
In the following, we shall discuss what happens when both are introduced.

Note that the theory we have obtained so far has a closed operator algebra,
and does not involve indecomposable modules. We should therefore not expect
to encounter logarithms in computations of correlators.

\section{Logarithmic lift of the $\su(2)_{-1/2}$ model}

\subsection{Symplectic fermions}

Heuristically, the extension from the small to the large algebra above is done
by keeping the states which were removed by the cohomology of
$\xi_0$. As already mentioned, we could have studied the situation
when keeping states removed by the cohomology
of its companion instead.
Here we want to consider the situation where no states are removed by
the cohomology of a BRST operator. This leads to the introduction
of  `symplectic fermions', $\psi_1$ and $\psi_2$, satisfying
\cite {Kau}
\ben
        \psi_1(z)\psi_2(w) = - \ln (z-w)+:\psi_1(z)\psi_2(w):\;.
\label{sf}
\een
The extension is first done by integrating
\ben
        \pa\psi_2 = \eta\;,
\een
resulting in the mode expansion
\ben
        \psi_2(z) = \alpha_2 + \eta_0 \log z - \sum_{n\in \ZZ\setminus \{0\}}
         \frac{1}{n} \eta_n z^{-n}\;,
\een
thereby introducing an extra zero mode $\alpha_2$.  This leads
to the conjugate field $\psi_1$ being a deformation of $\xi$
\ben
        \psi_1 = \xi + \al_1 \log z\;,
\een
with $\{\al_2,\al_1\}=1$ and
\ben
        \psi_1(z)=\xi_0+\al_1\log z+\sum_{n\in \ZZ\setminus \{0\}}
    \xi_n z^{-n}\;.
\een
One observes that when restricting to the small algebra,
generated here by $\pa\psi_1$ and $\pa\psi_2$,
one has the same algebra and irreducible modules as we found in the
first part of this paper.

The energy-momentum tensor is constructed by using the deformation
performed in \cite{FFHST}\footnote{A general study of how certain
indecomposable
modules may be obtained by considering short exact sequences involving
extensions of ordinary modules, may be found in \cite{FFHST}.
Their construction is not directly related to our results below, though.},
or directly using the symplectic fermion construction:
\ben
        \tilde{T}(z)= \pa\xi \eta(z)
         +\frac{\al_1 \eta(z)}{z}=\pa\psi_1\pa\psi_2(z)\;.
\een
The full energy-momentum tensor then reads
\ben
        T=\hf\pa\phi\pa\phi+\pa\psi_1\pa\psi_2\;.
\label{Tsf}
\een
Already at this level, we observe the vacuum Jordan cell
($\eta_0\ket{\Omega}=\al_1\ket{\Omega}=0$),
\ben
        \tilde{L}_0 \alpha_2 \xi_0\ket{\Omega}
         = \ket{\Omega}, \ \ \ \ \tilde{L}_0\ket{\Omega}=0\;.
\een
Similarly, the generators of the $\su(2)$ symmetry become
to
\bea
        J^+ &=& \hf e^{-2i\phi} \pa^2\psi_2 \pa\psi_2 \;,\nn
        J^3 &=& \frac{i}{2}\pa\phi\;,\nn
        J^- &=& -\hf e^{2i\phi} \pa^2\psi_1 \pa\psi_1\;.
\eea
They still satisfy the same algebra, albeit the space on which
they act is augmented by zero modes.  In terms of the old variables,
the above representation leads to the deformation of $J^-$ only,
and we have
\ben
        J^- = \frac{1}{2} e^{2i\phi} \left(
         \pa^2\xi\pa\xi -\frac{\al_1\pa\xi}{z^2}
         - \frac{\al_1\pa^2\xi}{z}\right)\;.
\een
Here we can describe the variable $\alpha_2$ as the one enabling
the creation of the new indecomposable modules.  Essentially, the introduction
of the zero mode creates an auxiliary space in which the larger affine
representations are embedded. The zero mode $\al_1$, on the other hand,
acts trivially on the states not containing $\alpha_2$
(i.e.,  $\al_1\ket{\Omega}=0$
for example).  On the states containing $\alpha_2$ it allows the transition
between the states in the auxiliary space.

It should be noted that the deformation presented above is not different
from the one found in $c=-2$ theories, and therefore the results found
there apply here as well.
More interesting results are found when one studies the $su(2)$
indecomposable representations. This is what is done in the next sections.

The presence of indecomposable representations of $\su(2)$ results
in a logarithmic conformal field theory.

\subsection{Indecomposable $j=0$ module}

Let us analyze the module generated from $\ket{\omega}$, the
Jordan-cell partner to the vacuum state $\ket{\Omega}$. The corresponding
fields are
\ben
        \ket{\omega}\lra\omega=\psi_2\psi_1,\ \ \ \ \ \ \
         \ket{\Omega}\lra\Omega=1\;.
\een
{}From the OPE
\ben
       T(z)\omega(w)= \frac{1}{(z-w)^2}+ \frac{\pa\omega(w)}{(z-w)}
        +{\cal O}((z-w)^0)\;,
\een
we immediately recover the Jordan-cell structure
$L_0\ket{\omega}=\ket{\Omega}$.
Due to the non-diagonal action of $L_0$, the
module generated from $\ket{\omega}$ is indecomposable. We shall
denote it $\IM_0$ (where the $\IM$ refers to it being indecomposable).

To unravel the detailed structure of the module $\IM_0$, we will evaluate
various products of the currents with the field $\omega$. Consider first
\bea
        J^+(z)\omega(w)&=&\hf \left(e^{-2i\phi} \pa^2
         \psi_2\pa\psi_2\right)(z)\,(\psi_2\psi_1)(w) \nn
        &=&\frac{1}{(z-w)^2}\left( -\frac{1}{2} e^{-2i\phi}
         \pa\psi_2\psi_2\right)\nn
        &+&\frac{1}{z-w}\left(-e^{-2i\phi}
         \pa^2\psi_2\psi_2+i\pa\phi e^{-2i\phi}\pa\psi_2\psi_2\right)\nn
        &+&(z-w)^0\left(\hf e^{-2i\phi}\pa^2\psi_2\pa\psi_2\psi_2\psi_1
         -\frac{3}{4}e^{-2i\phi}\pa^3\psi_2\psi_2\right.\nn
        &&\left.+2i\pa\phi e^{-2i\phi}\pa^2\psi_2\psi_2
         +\frac{i}{2}\pa^2\phi e^{-2i\phi}\pa\psi_2\psi_2
         +(\pa\phi)^2e^{-2i\phi}\pa\psi_2\psi_2\right)\nn
        &+&{\cal O}(z-w)\;.
\label{J+omega}
\eea
In the final expression, all arguments are evaluated at  $w$.
{}From this expansion, we read off the action of the various modes
$J^+_{n\geq-1}$ on $\omega$ as
\bea
        J^+_{n\geq2}\omega&=&0\;,\nn
        J^+_1\omega&=&-\hf e^{-2i\phi}\pa\psi_2\psi_2\;,\nn
        J^+_0\omega&=&-e^{-2i\phi}\pa^2\psi_2\psi_2
         +i\pa\phi e^{-2i\phi}\pa\psi_2\psi_2\;,\nn
        J^+_{-1}\omega&=&\hf e^{-2i\phi}\pa^2\psi_2\pa\psi_2\psi_2\psi_1
         -\frac{3}{4}e^{-2i\phi}\pa^3\psi_2\psi_2
         +2i\pa\phi e^{-2i\phi}\pa^2\psi_2\psi_2\nn
        &&+\frac{i}{2}\pa^2\phi e^{-2i\phi}\pa\psi_2\psi_2
         +(\pa\phi)^2e^{-2i\phi}\pa\psi_2\psi_2 \;.
\label{J+nomega}
\eea
The first observation to be made from these computations is that
$|\omega\R$ is not an affine highest-weight state
since $ J^+_1\omega\not=0$ (even though it is a Virasoro
highest-weight state). Moreover, the action of the various modes
$J^+_{n}$ on $\omega$
do not produce fields that belong to the identity module: all the
fields generated so far contain a $\psi_2$ without derivatives.

\begin{figure}
\centerline{\epsfig{figure=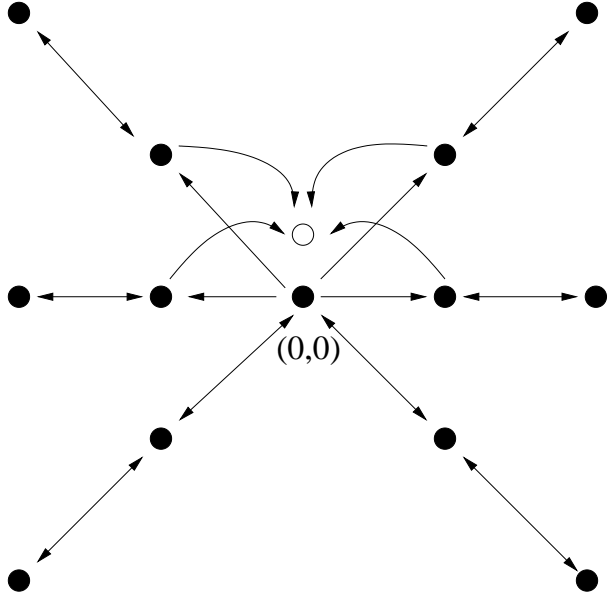,width=8cm}}
\vspace{0.3cm}
\caption{The indecomposable $j=0$ module $\IM_0$.}
\label{FigRep0}
\end{figure}

The extremal diagram of the module $\IM_0$ is displayed  in
Fig. \ref{FigRep0}. The central black dot is associated to the
$\omega$ field. So far,
we have described the three arrows directed toward the right  that
leave the point $\omega$, and these all connect other black dots.
A black dot in Fig. \ref{FigRep0} is always associated to a field
containing either $\psi_1$ or $\psi_2$ (or both) without derivatives.
All other fields in the module are represented by open dots.

Consider now the field $J^+_1\omega$. Acting on it with $J^+(z)$ gives
\ben
       J^+(z)(J^+_1\omega)(w)=
        \frac{1}{(z-w)^2}\left(-\frac{1}{8}
        e^{-4i\phi}\pa^3\psi_2\pa^2\psi_2\pa\psi_2\psi_2\right)(w)
        +{\cal O}((z-w)^{-1})\;,
\een
which shows that $(J^+_1)^2\omega\not=0$. More generally, we find that
\ben
       (J^+_1)^{n+1}\omega=\left(\frac{-1}{2^{n+1}\prod_{m=1}^{2n}m!}\right)
        e^{-2(n+1)i\phi}\pa^{2n+1}\psi_2\dots\pa^2\psi_2\pa\psi_2\psi_2\;.
\een
That $(J^+_1)^2\omega$ is non-vanishing accounts for the two
N-E arrows in Fig. \ref{FigRep0}. Consider now the action of $J^-(z)$
on the field $(J^+_1)^2\omega$:
\ben
       J^-(z)(J^+_1)^2\omega(w)=
        \left(3e^{-2i\phi}\pa\psi_2\psi_2\right)(w)+{\cal O}(z-w)\;.
\een
The absence of a single pole implies that
$J^-_0(J^+_1)^2\omega=0$. From the first regular term, we infer that
\ben
       J^-_{-1}(J^+_1)^2\omega\ \propto\ J^+_1\omega\;.
\een
We have thus a double-sided arrow linking $J^+_1\omega$ and
$(J^+_1)^2\omega$ in  Fig. \ref{FigRep0}.

In the same way, let us consider the action of $J^-(z)$ on
$(J^+_1\omega)(w)$:
\ben
       J^-(z)(J^+_1\omega)(w)= (z-w)^0 (1)+{\cal O}(z-w)\;.
\een
This readily implies that $J^-_0J^+_1\omega=0$. But more
importantly, this OPE shows
that the action of $J^-_{-1}$ on $J^+_1\omega$ is not proportional to
$\omega$. Instead, it is proportional  to the identity field.
This reflects the indecomposability of the module.
The arrow linking $\omega$ to $J^+_1\omega$ is thus not double-sided.
There is an S-W arrow leaving $J^+_1\omega$ but it links it to the open dot
representing the identity field, associated to $|\Omega\R$.
The vectors in the module generated from $|\Omega \R$
(the highest-weight vector in the irreducible module $\VMR_0$)
may be thought of as occupying integer (charge, level)
lattice points ($(m,h)$ with $h\geq |m|$) in a separate but
equivalent layer underneath the layer limited by black dots in
the downward part of the extremal diagram.

The other arrows in  Fig. \ref{FigRep0} are obtained analogously.
We note the similarity of this module with the extended $j=0$ module
in the $\su(2)_{-4/3}$ WZW model discussed in
\cite{Gab} (there denoted ${\cal R}_0$).\footnote{There is a minor difference
in the picture provided in \cite{Gab}, as the $L_0$ direction is inverted
in that paper.}

As indicated in Fig. \ref{FigRep0}, the vectors $J^+_1\ket{\omega}$ and
$J^-_1\ket{\omega}$ each generate a submodule in $\IM_0$.
Focusing on the module generated from
the vector $J^+_1\ket{\omega}$, we immediately see that it corresponds to
a $\pi_1$ twist of a module similar to the relaxed
module with a charged singular vector in Fig. \ref{large2}.
Similarly, the module generated from the vector $J^-_1\ket{\omega}$
corresponds to a $\pi_{-1}$ twist of a module similar to the relaxed
module with a charged singular vector in Fig. \ref{large1}.
Note that both modules have the irreducible module $\VMR_0$, generated from
$\ket{\Omega}$, as a submodule. This is simply due to the single-sided
arrow linking $J_1^\pm\omega$ to $|\Omega\R$.

\subsection{The $\IM_{0}$ character}

As discussed above, acting on the states $J_1^+\omega$ and $J_1^-\omega$
generates submodules that are twists of relaxed modules
containing charged singular vectors, i.e.,
$\psi_2\partial\psi_2 e^{-2i\phi} =
\pi_1(\psi_2\partial\psi_2 e^{-3i\phi/2})$ and $\psi_1 \partial\psi_1
e^{2i\phi}=\pi_{-1}(\psi_1 \partial\psi_1 e^{3i\phi/2})$.
Since these states generate relaxed highest-weight modules in
which the charged singular vector generates exactly the identity module, their
irreducible module\footnote{Here we really consider the irreducible modules
$\RMR$ from which the charged singular vector has been factored out.}
$\chi^{\RMR}_{\mu_1,\mu_2;\theta}(z,q)$ will not count the identity module.
In turn, the full indecomposable module will count each state in the identity
module twice. It is not difficult to see that we have a character of the form
\bea
     \chi^\IM_{j=0}(z,q)&=&2\chi_{0;0}(z,q) + \chi_{-1,\frac{3}{2};1}(z,q) +
      \chi_{0,\frac{1}{2};-1}(z,q)\nn
     &=&\chi^{R^a}_{-1,\frac{3}{2};1}(z,q) +
      \chi^{R^a}_{0,\frac{1}{2};-1}(z,q)\;,
\label{cIj0}
\eea
where each character is understood as a formal series.
This appears to be a new observation.

\subsection{Indecomposable $j=1/2$ module}

Here we demonstrate how the symplectic fermions admit an indecomposable
extension, $\IM_{1/2}$, of the irreducible $\su(2)_{-1/2}$ spin-1/2 module
$\VMR_{1/2}$. Proceeding in a way analogous to the construction of the
$\IM_0$ module above, we first need
to identify the candidate Jordan-cell partners to the $\beta$ and
$\gamma$ fields. We recall that the two ghost fields constitute
an $su(2)$ spin-1/2 representation. It is natural to expect
that each of these fields is to be combined with
its $\omega$-composite. We first focus on $\beta$, in which case
we thus expect the pair $\beta$ and
\ben
       \beta\omega= e^{-i\phi}\pa\psi_2 \psi_2 \psi_1\;,
\een
to form a Jordan cell. Let us check this by computing the
OPE of $T$ with $\beta\omega$:
\ben
       T(z)\beta\omega(w) = \frac{(-e^{-i\phi}\psi_2)(w)}{(z-w)^3}+
        \frac{(\beta+\frac{1}{2}\beta\omega)(w)}{(z-w)^2}
        +\frac{(\pa(\beta\omega))(w)}{(z-w)}+{\cal O}((z-w)^0)\;.
\een
This shows that our candidate field $\beta\omega$ is indeed the
Jordan-cell partner to $\beta$ since
\ben
       L_0|\beta\omega\R = \frac{1}{2}|\beta\omega\R  + |\beta\R\;,
\een
as required. The above OPE also shows that $\beta\omega$ is not a
Virasoro primary field, since
\ben
       L_1\beta\omega = -e^{-i\phi}\psi_2\;.
\een

\begin{figure}
\centerline{\epsfig{figure=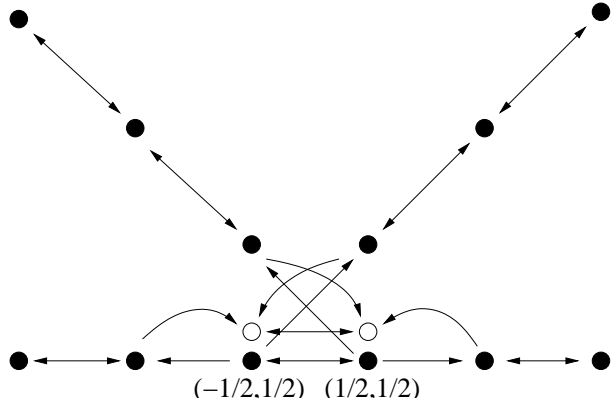,width=8cm}}
\vspace{0.3cm}
\caption{The indecomposable $j=1/2$ module $\IM_{1/2}$.}
\label{FigRep1/2}
\end{figure}

Let us analyze the affine algebraic structure of the module
built from the state
$|\beta\omega\R$. Consider first the action of the current $J^+(z)$ on
$\beta\omega(w)$. Extracting the contribution of the first three
non-vanishing terms of the OPE  yields
\bea
       J^+_{2}\beta\omega&=&e^{-3i\phi}\pa^2\psi_2\psi_2  \;, \nn
        J^+_1\beta\omega&=& e^{-3i\phi}[-2i\pa\phi\pa^2\psi_2\pa\psi_2\psi_2+3
        \pa^3\psi_2\pa\psi_2\psi_2] \;,\nn
       J^+_0\beta\omega&=&e^{-3i\phi}\left[
        \frac{1}{3}\pa^4\psi_2\pa\psi_2\psi_2-\frac{3}{2}
         i\pa\phi\pa^3\psi_2\pa\psi_2\psi_2 \right.\nn &&\left.
        \qquad\phantom{\frac{1}{1}}+\, \pa^2\psi_2
        \pa\psi_2\psi_2[-2(\pa\phi)^2-i\pa^2\phi]\right]\;.
\label{J+betanomega}
\eea
This computation unravels a part of the indecomposable module
$\IM_{1/2}$ displayed in Fig. \ref{FigRep1/2}.
Again, the black dots are associated to the new
fields that appear upon extending the model by introducing the
symplectic fermions (as opposed to working in the small algebra).
The $\beta\omega$ field itself is the black dot
labeled $(1/2,1/2)$. We thus have arrows
pointing in the N-E direction, one by two units (corresponding to
$J^+_{2}\beta\omega$ which lies on the extremal diagram), and one by
one unit (for $J^+_{1}\beta\omega$, not drawn),
as well as an arrow pointing horizontally to the right
($J^+_{0}\beta\omega$). The resulting states all contain a symplectic
fermion on which no derivative acts. The reached dots are thus all black.

Consider now the action of the current $J^-(z)$ on the three fields
just described. At first we have
\bea
     J^-(z)J^+_{2}\beta\omega(w)&=&\left(e^{-i\phi}\psi_2\right)(w)
      +(z-w)\left(e^{-i\phi}[2i\pa\phi\psi_2-2\pa\psi_2]\right)(w)\nn
     &+&{\cal O}((z-w)^2)\;.
\eea
That $J^-_0J^+_{2}\beta\omega=0$ confirms that the oblique line is
really extremal. The $J^-_{-1}$ mode produces the field
\ben
       J^-_{-1}J^+_{2}\beta\omega= e^{-i\phi}\psi_2\;,
\een
which has previously been encountered as $L_1 \beta\omega$.  It is also easily
checked to be proportional to $J_{1}^+\gamma\omega$. Considering the
next term in the expansion, we find that
$J^-_{-2}J^+_{2}\beta\omega$ does not return to
$\beta\omega$; it is instead expressed as a sum of two fields:
\ben
       J^-_{-2}J^+_{2}\beta\omega= 2ie^{-i\phi}\pa\phi\psi_2-
2e^{-i\phi}\pa\psi_2\;.
\een
The first one is a descendant of $J^-_{-1}J^+_{2}\beta\omega$. The
other one is proportional to $\beta$.
That confirms the indecomposable character of the representation.

We perform a similar computation on the field $J^+_{1}\beta\omega$
and find that
\ben
       J^-_{0}J^+_{1}\beta\omega\ \propto\ e^{-i\phi}\psi_2\;,
\een
while $ J^-_{-1}J^+_{1}\beta\omega $ is a linear combination of $\beta$ and
a descendant of $e^{-i\phi}\psi_2 $.
The corresponding arrows are not drawn since they are not in the extremal
part of the diagram. We also find that
\ben
       J^-_{1}J^+_{0}\beta\omega\ \propto\ e^{-i\phi}\psi_2\;,
\een
and again $J^-_{0}J^+_{0}\beta\omega$ is  a linear combination of $ \beta$ and
$J^3_{-1}e^{-i\phi}\psi_2$.
This shows that the arrow pointing from $\beta\omega$ to
$J^+_{0}\beta\omega$ is not
two-sided, as indicated. The arrow leaving $J^+_{0}\beta\omega$ ends
on the open dot representing $\beta$.

The rest of the extremal diagram is completed straightforwardly.
Note that the two new fields
$\beta\omega$ and $\gamma\omega$ form a sort of $su(2)$ doublet,
albeit immersed into an extended module,
\ben
        J_0^-\beta\omega\ \propto\ \gamma\omega+\cdots \quad,\qquad
        J_0^+\gamma\omega\ \propto\ \beta\omega+\cdots\;,
\een
where the dots refer to a descendant of the state just
above $\gamma\omega$ or $\beta\omega$, respectively.

\subsection{Closure of the fusion algebra}

Twisting the two indecomposable modules just obtained will produce all other
indecomposable modules that extend the regular spectrally flowed modules.
In particular, we find $\IM_{-1/4}$ and  $\IM_{-3/4}$ by applying
$\pi_{-1}$ to $\IM_{0}$ and $\IM_{1/2}$, respectively. This indicates
that to every $h=-1/8$ twist field associated to a state at level zero
in a highest- or lowest-weight representation, namely $\tau_n$ with
$n\in\ZZ$ (see \cite{LMRS}, for example)
\ben
       \tau_n= \sigma_ne^{i(n-1/2)\phi}\;,
\een
where
\ben
       \sigma_n =  \pa\psi_1 \cdots \pa^{n-1}\psi_1 \,,\qquad
        \sigma_{-n}= \pa\psi_2 \cdots \pa^{n-1}\psi_2\;,
\een
we can associate the Jordan-cell partner $\tau_n\omega$.
For instance, $\tau_1=e^{i\phi/2}$
is Jordan-paired with $\tau_1\omega=e^{i\phi/2}\psi_2\psi_1$. This
is easily confirmed by an
explicit computation of the OPE of $T$ with $e^{i\phi/2}\psi_2\psi_1$.

At this point in our analysis of the lifted theory,
we have found that all  admissible representations and their spectral flows
have indecomposable Jordan-cell partners.
It remains to be understood if the (almost reduced) relaxed modules
also have indecomposable Jordan-cell partners.
The resolution is obvious, as we have already observed that the
relaxed modules correspond to composite fields involving exactly one
of the symplectic fermions, $\psi_1$ or $\psi_2$, without derivatives.
On the other hand, the indecomposable structure is based on
Jordan-cell pairs $\Psi\omega$ and $\Psi$:
\ben
       L_0\ket{\Psi\omega}=h(\ket{\Psi\omega})+\ket{\Psi}\;,
\een
where $h$ denotes the conformal weight.
Since a symplectic fermion squares to zero, $\Psi$ cannot contain
an underived $\psi_i$, and hence it cannot correspond to an element
in a relaxed module. We may therefore conclude that {\sl there
are no Jordan-cell partners to the relaxed modules}.
By going from the large algebra to the symplectic formalism,
we simply double the number of relaxed modules (see the comment on
asymmetry at the end of section 4) and introduce indecomposable
representations as Jordan-cell partners to the (ordinary) admissible
representations and their twists.

A crucial  step toward determining if we now have a complete description
of the field content of this $\su(2)_{-1/2}$ logarithmic conformal field
theory, is to verify that the fusion algebra closes. But closure is an
immediate consequence of our free-field construction for the indecomposable
representations, and of the closure of the small and large
algebra representations.

\section{The partition function}

Using the free-field representation, we have found that each version
of the $\su(2)_{-1/2}$ WZW model (based on the small, large or symplectic
formulation of the constituent $c=-2$ theory)
has a closed fusion algebra. However, this does not automatically ensure
that all the fields of the physical theory are accounted for. To address this
question, it is traditional in the case of rational CFTs to
consider  modular invariant partition functions. We try to apply this
approach to our problem  here, and encounter considerable difficulties.

The  basic result underlying most efforts to
make sense of WZW models at
fractional level is the observation by Kac and Wakimoto \cite{KW}
that for a given admissible level,
there is a finite number of admissible representations, that transform
linearly among themselves under modular transformations.
This readily leads to a formal modular invariant
which is simply the diagonal invariant built out of the admissible
{\it character functions} \cite{FSST} $F_{j}(z,q)$ (which are the
Kac-Wakimoto admissible characters \cite{KW} -- whose relationship to the
admissible characters $\chi_j$ that correspond to genuine power series are
reviewed in sect. 2.4). For our case, we are only interested in this
diagonal modular invariant\footnote{The
$\su(2)_k$ modular invariants have been classified in \cite{Lu}.}:
\ben
\label{modinn}
M(z,q)=|F_{0}(z,q)|^2+|F_{\frac{1}{2}}(z,q)|^2+|F_{-\frac{1}{4}}(z,q)|^2+
         |F_{-\frac{3}{4}}(z,q)|^2\;.
\een
A natural but complex question is the relation of this expression
with the various
incarnations of the $\beta\gamma$ system  which we have discussed in this and
our previous paper \cite{LMRS}.

As explained before, the four
character functions are mapped onto each other under the spectral
flow.
Forget for a moment the subtle difference (due to convergence issues) between
these character functions and the characters proper.
The partition function, defined as usual as  the trace over the
whole space of states of $q^{L_0-c/24}\bar{q}^{\bar{L}_0-c/24}
z^{J_0^3}\bar{z}^{\bar{J}_0^3}$, for the `small algebra'
$\beta\gamma$ system  should then read formally
\ben
      Z=\left(\sum_{\theta=-\infty}^{\infty} 1\right)
      \left\{|F_{0}(z,q)|^{2}+|F_{1/2}(z,q)|^{2}+|F_{-1/4}(z,q)|^{2}+
       |F_{-3/4}(z,q)|^{2}\right\}\;,
\label{modinv}
\een
where the infinite sum accounts for the sum over all the spectrally
flowed representations. As  far as modular
invariance is concerned, this sum, since it factors out as a
numerical prefactor, is  irrelevant,  and
     one simply recovers
the original Kac and Wakimoto result. Physically, the
appearance of the
prefactor is a `reminder' of the divergence of the initial
$\beta\gamma$
functional integral, or the unbounded (from below) nature of the
spectrum. Actually,,
using the character functions instead of the true characters has
provided a regularization of this  problem.

We now discuss more the issue of characters versus character
functions, which is due
to `contact terms'. The key observation is that
\ben
      {1\over 1-zq^{a}}\ =\ \left\{ \begin{array}{ll}
       \phantom{-}\sum_{0}^{\infty}z^{n}q^{na},~~~\hspace{1cm}\
|zq^{a}|<1\;,\\
\mbox{}\\
       -\sum_{1}^{\infty}z^{-n}q^{-na},~~~\ \ \ \ |zq^{a}|>1\;.
	\end{array} \right.
\label{basic}
\een
Therefore, inside the circle of convergence, the  series coincides
        with the analytic expression
        \begin{equation}
	\hbox{character}=\sum_{0}^{\infty}z^{n}q^{na}=\hbox{
	character function},~~~ |zq^{a}|<1\nonumber\;,\\
	\end{equation}
while outside, they differ. The character is still defined by the
        same formal sum, and thus
        \begin{eqnarray}
	\hbox{character}=\sum_{0}^{\infty}z^{n}q^{na}&=&\hbox{
	character function}+\sum_{-\infty}^{\infty}
	z^{n}q^{na}\nonumber\\
	&=&
	\hbox{ character function}+\delta(zq^{a},1),~~~ |zq^{a}|>1\;.
	\end{eqnarray}

The contact terms allow one to express the  characters in terms
of the character functions. One finds first
\begin{eqnarray} \label{basic2}
        \chi_{j(1,1),{3\over 2};0}(z,q) &=&F_{0}(z,q)
     +q^{1/24}
       \prod_{1}^{\infty}{1\over (1-q^{n})^{2}} \sum_{0}^{N}
      (-1)^{p}q^{p(p+1)/2}\delta(z
        q^{2p+1},1)\;,\nonumber\\
\chi_{j(2,1),{3\over 2};0}(z,q) &=&  F_{1/2}(z,q)+q^{1/24}
       \prod_{1}^{\infty}{1\over (1-q^{n})^{2}}
\sum_{-M}^{-1}(-1)^{p}q^{p(p+1)/2}\delta(z
        q^{2p+1},1)\;,
        \end{eqnarray}
where the bounds are as follows.  Either $z$ is outside
the first disk of convergence, in which case the lower bound is $0$
and the higher bound $N\geq 0$ follows from $|q|^{-2N-1}<|z|<|q|^{-2N-3}$;
or $z$ is inside the first disk, and then $M\geq 1$ is obtained from
$|q|^{2M+1}<|z|<|q|^{2M-1}$.

A similar but different expression holds for the other characters
\begin{eqnarray}\label{basicI}
        \chi_{j(2,2),{3\over 2};0}(z,q)
         &=&F_{-1/4}(z,q)
         +q^{-1/12}
        \prod_{1}^{\infty}{1\over (1-q^{n})^{2}}
        \sum_{1}^{N}
      (-1)^{p}q^{p^{2}/2}\delta(z
        q^{2p},1)\;,\nonumber\\
        \chi_{j(1,2),{3\over 2};0}(z,q) &=& F_{-3/4}(z,q)
+q^{-1/12}
        \prod_{1}^{\infty}{1\over (1-q^{n})^{2}}
        \sum_{-M}^{0}
      (-1)^{p}q^{p^{2}/2}\delta(z
        q^{2p},1)\;.
\end{eqnarray}
Here, either $z$ is outside
        the first disk of convergence, in which case the lower bound is
        $1$
        and the higher bound $N\geq 0$ follows from
        $|q|^{-2N}<|z|<|q|^{-2N-2}$;
        or $z$ is inside the first disk, and then $M\geq 0$ is obtained from
        $|q|^{2M+2}<|z|<|q|^{2M}$.

The operator content of
the theory includes all the representations obtained by the action of
the spectral flow. In some cases, the result is a representation with
the spectrum of $L_{0}$ bounded from below, but in most cases it is a
representation with the spectrum not bounded from below.  This
spectrum can be deduced from
\begin{equation}
         \chi_{j,{3\over 2};\theta}(z,q)=
         q^{-\theta^{2}/8}z^{-\theta/4}\chi_{j,{3\over
2};0}(q,zq^{\theta})\;.
         \end{equation}
It so
happens however that the contribution of the unbounded part of the
spectrum is essentially encoded into contact terms (and as a result,
the
character functions map nicely onto each other under the action of
the flow). For instance, inside the domain $|q|<z<|q|^{-1}$,
the character function $F_{0}$ coincides with the  character
of the identity $\chi_{0,{3\over 2};0}$. Inside the next domain
$|q|^{-1}<|z|<|q|^{-3}$, the character function instead coincides
with the  spectral flowed object $-\chi_{{1\over 2},{3\over 2};2}$, since,
in the domain of interest, there are
no contact terms , that is
\begin{equation}
        \chi_{{1\over 2},{3\over 2};\theta=2}=F_{1/2,\theta=2}=
        -F_{0},~~~|q|^{-1}<|z|<|q|^{-3}\;,
	\end{equation}
and so on.

This suggests another way to define the partition function: one may
relax the requirement that
`each operator' is counted for any value of $q,z$, and instead
assume that the operator content
is obtained by patching the spectra obtained in each domain of
convergence (this idea is also suggested in \cite{SS},\cite{FSST}).
For instance, the term
$|F_0|^2$ would be interpreted as encoding the
     untwisted $j=0$ representation in the first annulus, {\sl plus}
     the $j=1/2,w=2$ twisted  representation in the next annulus, etc. At
the moment, we have no clear physical  justification for this
approach, but it seems worth investigating.
One then gets the partition function
\ben
\label{fausse}
     {Z}(z,q)= \sum_{j=0,1/2,\atop -1/4,-3/4} \sum_{\theta\in 2\ZZ}
      |\chi_{j,\frac{3}{2};\theta}(z,q)|^2ch_{j,{3\over 2};\theta}\;,
\label{FausseFonction}
\een
where $ch$ is a characteristic function, equal to unity in
the annulus of convergence, and $0$ otherwise. This almost
gives the Kac Wakimoto invariant without the infinite
     multiplicity prefactor.

The difficulty here arises from the fact that expression
(\ref{fausse}) provides a covering of the complex $z$ plane
(assuming $q$ to be fixed with $|q|<1$) but only up to the boundaries
between the annuli since the region of convergence is given by strict
inequalities. One thus realizes that in order to have a full
covering of the complex $z$  plane, something is
missing: the circles at the boundaries of the different
annuli, which are in fact necessary to
recover the Kac-Wakimoto invariant.

To get a clearer picture, let us pause to study the effect of modular
transformations
on the circles of convergence and the poles.
Write
\bea
     \tau\, \rightarrow\,  {a\tau +b\over c\tau +d}\; , \qquad \zeta\,
      \rightarrow \, {\zeta\over  c\tau +d}\qquad\qquad (ad-bc=1)\;,
\eea
where $q=e^{2\pi i\tau}$ and $z= e^{2\pi i \zeta}$.
Let us start from a point $z$  such that $z=q^n$, $n\in\ZZ$, which is a
pole for half of the (flowed) admissible representations. It means
that $\zeta = n\tau+m$ for some integer $m$.
Now consider the modular transformation
\bea
     z\,\rightarrow\, z'= e^{2\pi i \, {\zeta\over  c\tau +d}}=  e^{2\pi i
      \, {n\tau+m\over  c\tau +d}}\;.
\eea
To write $z'$ in the form $z'= (q')^N $ for $ N\in \ZZ$ requires
\bea
     {n\tau+m\over  c\tau +d} = N\, {a\tau +b\over c\tau +d} +M\;,
\eea
for some $M$. Solving for $N$ and $M$, using $ad-bc=1$ yields
\bea
     N=nd-mc\;,\qquad M=ma-nb\;,
\eea
and these are indeed both integers. We thus fall back on another
point that is also a pole for half of the admissible representations.
Note, however, that $N$ does not necessarily have the same parity as $n$.
(Recall that when $s=1$, poles appear for $n$ odd, while for $s=2$
poles correspond to $n$ even). However, under modular transformation, the
fields transform linearly in terms of all the other fields. In
particular, the $S$ matrix for this model is
\ben
     S={1\over 2} \pmatrix{\phantom{-}1&-1&-1&\phantom{-}1\cr
      -1&\phantom{-}1&-1&\phantom{-}1\cr
      -1&-1&-i&-i\cr \phantom{-}1&\phantom{-}1&-i&-i\cr}\;,
\een
in the order $j=0,1/2,-3/4,-1/4$. For instance, $F_0$ transforms as
\ben
     F_0(-1/\tau, \zeta/\tau) = {1\over 2}
      (F_0(\tau,\zeta)-F_{1/2}(\tau,\zeta)
      -F_{-3/4}(\tau,\zeta)+F_{-1/4}(\tau,\zeta))\;.
\een
On any circle, half of these representations have poles. That means
that regular points can be mapped to poles
and vice-versa. So either all the circles in consideration
(even the regular ones) are omitted,  or all the circles, and thus
all the poles, need to be included.

It is a bit  uncomfortable mathematically to exclude all the circles
from the definition of the partition function. Also, while we had
little rationale to begin with to justify the patching of the regions
of convergence in the partition function, it seems even more
arbitrary to exclude circles where the expressions converge. The only
tenable choice seems thus to define the partition function as the
Kac-Wakimoto invariant, and try to  explain, within this `patching
philosophy' the added circles,
including the poles.

The simplest explanation could be that the circles have to be there
to make $Z$ well defined mathematically, and do not have a physical
meaning. A more complex explanation would be that the
circles somehow encode the relaxed representations. While the fact that
characters of these representations appear as residues is a
tentalizing suggestion in that direction, we have not been able to
propose a consistent picture compatible with this interpretation.
Were we able to, this would probably show that the $\beta\gamma$
system based on the `small algebra' is not a consistent theory on
the torus, and that the extension to the `large algebra' is
necessary.

A last direction of attack would consist in manipulating formal
series. Notice indeed that,  using
\begin{equation}
        \chi_{0,{3\over 2};0}=F_{0}+q^{1/24}
     \prod_{1}^{\infty}{1\over (1-q^{n})^{2}}
\delta(zq,1),~~~|q|^{-1}<|z|<|q|^{-3}\;,
           \end{equation}
we have
\begin{equation}
        \chi_{{1\over 2},{3\over 2};2}=-\chi_{0,{3\over 2};0}+q^{1/24}
           \prod_{1}^{\infty}{1\over (1-q^{n})^{2}} \delta(zq,1)\;,
           \end{equation}
an expression which is now valid irrespective of the values of $z,q$.
The argument generalizes easily. One finds in particular
\begin{equation}
        \chi_{{1\over 2},{3\over 2};-2}=-\chi_{0,{3\over 2};0}-q^{1/24}
       \prod_{1}^{\infty}{1\over (1-q^{n})^{2}} \delta(zq^{-1},1)\;,
       \end{equation}
while identical expressions hold by switching the  representations
$1/2,0$. One also has
\begin{eqnarray}
        \chi_{0,{3\over 2};1}&=&\chi_{-{1\over 4},{3\over 2};0}\;,\nonumber\\
        \chi_{0,{3\over 2};-1}&=&-\chi_{-{3\over 4},{3\over 2};0}+q^{-1/12}
           \prod_{1}^{\infty}{1\over (1-q^{n})^{2}} \delta(z,1)\;,
           \end{eqnarray}
and identical expressions obtained by switching $1/2,0$ and
    similarly
    $-1/4,-3/4$, etc.

It thus seems plausible that one can write the partition function as a
     sum of moduli squares of formal series (the characters) instead as
     the sum of meromorphic functions (the character functions) that we
     have tried using so far. The implementation of modular invariance
     and various other questions, however, do not make this approach
     significantly more fruitful than the others.

To conclude this section: while there are several ways to formally
relate the Kac-Wakimoto invariant to our analysis of the
$\beta\gamma$ system (including the necessary inclusion of the
spectral flow), a detailed analysis of the modular invariant
partition function has eluded us. In particular, we have not been
able to use the torus to distinguish between the `small' and `large'
algebra $\beta\gamma$ systems, let alone the `symplectic algebra'
based on the symplectic fermions. Maybe it is not possible.

\section{Conclusion}

In this work, we have extended our free-field representation analysis
of the non-unitary $\su(2)_{-1/2}$ WZW model \cite{LMRS}.
This representation uses
a $c=-2$ $\eta\xi$ ghost system and a Lorentzian boson. In \cite{LMRS} we
restricted ourselves to the small algebra. It was found that the
spectrum contains an infinite number of fields with arbitrarily large
negative dimensions. As a result, the model, in its small algebra
formulation, is not a rational CFT, being rather quasi-rational
\cite{MooreS} (i.e., there is an infinite but countable number of
primary fields, while only a finite number of
them appear in a given fusion).

Here we have considered a
two-step extension of our previous work: first to the large algebra, and
then to the symplectic description of the underlying $c=-2$ theory. At the
large algebra level, we have seen that new representations appear.
They are identified with some of the relaxed modules of \cite{FST, SS}.
The relevance of  relaxed modules in this
context appears to be a new observation.

The symplectic formulation, in turn, involves also indecomposable
representations. As demonstrated in some details, they appear in
simple OPE calculations based on our free-field representations.
The `emergence' of these indecomposable
representations is induced by the introduction of {\em two} zero modes
(those of the symplectic fermions $\xi$ and $\d^{-1}\eta$).
Our observation that the associated indecomposable characters
appear to be expressible as the sum of
almost reduced (relaxed) modules could potentially be important
and deserves a critical study.

At the symplectic level, we find that the overall
pattern mimics the one displayed in \cite{Gab} in the
$\su(2)_{-4/3}$ case, deduced there by somewhat
formal fusion-rule considerations.  Our results thus provide
independent, albeit implicit, support to the results of \cite{Gab}.

Our analysis has revealed a new example of a logarithmic
CFT, the logarithmic lift (i.e., the symplectic version)
of the $\su(2)_{-1/2}$ WZW model.
In turn, the explicit free-field representation offers a concrete
way of analyzing it. In fact, we believe that this model could serve
as an $c=-1$ counterpart to the paradigmatic $c=-2$ model
\cite{G}. The latter has been reviewed in much details in
\cite{Flohr} and \cite{Gaberdiel} where also vast lists of references
to other relevant works on $c=-2$ may be found.

Let us comment on the relation between these two models -- the
logarithmic $\su(2)_{-1/2}$ model and the $c=-2$ model.
It is natural that the logarithmic structure
observed in the $\su(2)_{-1/2}$ model must be inherited from that of
the $c=-2$ model. However, it should be stressed that it is not the
`original' logarithmic structure of the $c=-2$ model that shows up in
the $\su(2)_{-1/2}$ case.  The logarithmic solutions in the $c=-2$ model
were found through the analysis of the four-point correlation
function of the twist field $\sigma_{1/2}$, of dimension $-1/8$
\cite{G}.  This correlator involves two distinct channels
(intermediate states) with identical dimensions, leading to an explicit
logarithmic solution. That in turn induces a Jordan cell pattern.  But
this very twist field does not enter in our construction. Indeed,
we only use those twist fields of the
$c=-2$ model that can be expressed in terms of the $\eta\xi$ fermions. The
logarithmic structure (that is, the Jordan cells) arises solely due to the
presence of the two zero modes of the symplectic fermions $\xi$ and
$\d^{-1}\eta$, respectively. It would  be interesting
to understand yet other versions of the associated $\beta\gamma$ system
where twist fields in the $\eta\xi$ sector would now be involved.

In relation to the Jordan cells, let us remark that in the present
context we only observe the usual Virasoro Jordan cells:
\ben
  L_0 |\phi\rangle =  h |\phi\rangle + |\psi\rangle \; , \qquad
   L_0 |\psi\rangle =  h |\psi\rangle\ ,
\een
that is, we do not observe any Lie-type Jordan cell of the form
\ben
  J_0^3 |\phi\rangle =  m |\phi\rangle + |\psi\rangle \; , \qquad
   J_0^3 |\psi\rangle =  m |\psi\rangle\ .
\een
Whether these can occur in $\su(2)$ models remains an interesting
open question.

We have also discussed some issues related to the construction of a
modular invariant that would match the physical spectrum.  This has
turned out to be a rather tricky issue, on which no definite conclusions
have been presented.
In that vein, the symplectic version
of the $\su(2)_{-1/2}$ model may be the only well-defined one.
A further study of the characters and modular invariants
may help settling this important question.
Indeed, trying to unravel
this modular invariant problem in an unambiguous way is a natural
extension of this work. Another one is to relate the symplectic free-field
representation to the known logarithmic solutions to the KZ equations.

One could also ask to which extend our discussion of the $\su(2)_{-1/2}$
model applies to other admissible $\su(2)_{k}$ models.
As indicated in
section 2.3, the existence of  different layer-versions of the theory is
quite likely for all cases  where $2k$ is an (odd) integer. 
Recall that,
from a fusion-rule point of view, the models with $2k$ integer appear to
differ fundamentally from those with $2k$ non-integer. On the other
hand, for all (non-integer) admissible $k$, the top layer is
naturally expected to correspond to a logarithmic CFT.
This expectation stems from the analysis of two different fractional-level
$\su(2)_k$ WZW models (with $k=-4/3$ and $k=-1/2$, respectively)
based on two quite distinct approaches but with similar results.

Granting the validity of this general expectation, one can argue that these
admissible WZW models provide particularly good laboratory models to study
logarithmic CFTs. Indeed, they do not  embody unfamiliar types of
logarithmic structures (i.e., they have Virasoro and not Lie-type Jordan
cells) while their skeleton structure, namely the finitely many
admissible representations, encodes the crucial modular covariance
property.

Although admissible WZW models have yet to be fully tamed, it should be
stressed that their use as building blocks of non-unitary CFTs via the
diagonal cosets
\ben
{ {\widehat g}_k\oplus {\widehat g}_\ell\over {\widehat
g}_{k+\ell} }\; , \qquad \qquad (\ell\in \ZZ_+)\;,
\een
nevertheless seems exempt of any ambiguities:
complications are found to cancel out nicely between
the numerator and the denominator \cite{MSW}.
\vskip.5cm
%
%
\noindent {\bf Acknowledgment} We thank M. Flohr for helpful
correspondence. FL and JR thank T. Gannon for discussions, and PM
acknowledges a discussion with M. Gaberdiel.

\appendix

\section{Relaxed representations}

\subsection{Relaxed modules and charged singular vectors}

In this subsection, we introduce the concept of relaxed modules and
their twists following
\cite{FST,SS} (but adopting mainly the notation of the second reference).

Roughly, an untwisted relaxed module is a module defined on a state
satisfying the usual highest-weight conditions (the first line of eq.
(\ref{flow}) for $\theta=0$) except for the usual $su(2)$ hightest-weight
condition that the state is annihilated by $J^+_0$, which is not
enforced.  The term `relaxed' stems from the fact that one of the
defining conditions has been removed or relaxed.

More generally, for a fixed twist parameter, $\theta\in\ZZ$, the twisted
relaxed module $\RM_{\mu_1,\mu_2,t;\theta}$ is generated by the
operators $J^+_{n\leq\theta}$, $J^3_{n\leq-1}$,
and $J^-_{n\leq-\theta}$ acting on the vector
$\ket{\mu_1,\mu_2,t;\theta}$ satisfying the annihilation conditions
\ben
        J^+_{n\geq1+\theta}\ket{\mu_1,\mu_2,t;\theta}\ =\ J^3_{n\geq1}
         \ket{\mu_1,\mu_2,t;\theta}\ =\ J^-_{n\geq1-\theta}
         \ket{\mu_1,\mu_2,t;\theta}\ =\ 0\;,
\label{relc}
\een
and subject to
\bea
        J^-_{-\theta}J^+_\theta\ket{\mu_1,\mu_2,t;\theta}&=&-\mu_1\mu_2
         \ket{\mu_1,\mu_2,t;\theta}\;,\nn
        \left(J^3_0+\frac{k}{2}\theta\right)\ket{\mu_1,\mu_2,t;\theta}&=&
         -\hf(1+\mu_1+\mu_2)\ket{\mu_1,\mu_2,t;\theta}\;.
\eea
We recall that $t=k+2$. The Sugawara dimension, $\D_{\mu_1,\mu_2,t;\theta}$,
and charge, $m_{\mu_1,\mu_2,t;\theta}$, of the twisted relaxed highest-weight
vector $\ket{\mu_1,\mu_2,t;\theta}$ are
\bea
       &&\D_{\mu_1,\mu_2,t;\theta}=\frac{(\mu_1-\mu_2)^2-1}{4(k+2)}
        +\hf(1+\mu_1+\mu_2)\theta+\frac{k}{4}\theta^2\;,\nn
       &&m_{\mu_1,\mu_2,t;\theta}=-\hf(1+\mu_1+\mu_2)-\frac{k}{2}\theta\;.
\label{D}
\eea
Note that the construction is symmetric in $\mu_1$ and $\mu_2$.
When the twist parameter is omitted, it is understood to be zero and
the module or vector is not twisted. Again, comparing the conditions
(\ref{relc}) to the similar conditions defining the ordinary (twisted)
Verma module (\ref{verma}), we see that one condition has been
relaxed in the relaxed module.

A continuous representation (or rather module) of the finite-dimensional
Lie algebra $su(2)$ corresponds to the extremal
diagram of an untwisted relaxed module with neither $\mu_1$ nor
$\mu_2$ integer. It may be useful to recall here the notation
that is used in our previous paper \cite{LMRS} as well as in \cite{Gab}.
There continuous representations are denoted by $E_s$ and consist
of states $\ket{m}$ with
\begin{eqnarray}
     J_0^3\ket{m}&=&m\ket{m}\;,\nonumber\\
     J_0^+\ket{m}&=&\ket{m+1}\;,\nonumber\\
     J_0^-\ket{m}&=&\left(C-m(m-1)\right)\ket{m-1}\;,
\end{eqnarray}
with
\ben
C={(\mu_1-\mu_2)^2-1\over 4}\;,\qquad m\in\ZZ+s\;, \qquad
s=-{1\over 2}(1+\mu_1+\mu_2)\;.
\een The affine extension is obtained by acting
with the negative modes of the currents.

\begin{figure}
\centerline{\epsfig{figure=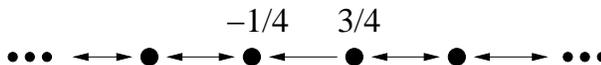,width=8cm}}
\vspace{0.3cm}
\caption{The extremal diagram of the relaxed module $\RM_{-1,-3/2,3/2}$
        with the charged singular vector $J^-_0\ket{-1,-3/2,3/2}$ with charge
        $m=-1/4$. Note that the left-most part of the diagram is the $su(2)$
      highest-weight representation with spin $j=-1/4$.}
\label{charged}
\end{figure}

Whenever $\mu_1$ is integer,
we have a so-called charged singular vector \cite{FST,SS} in the
extremal diagram, corresponding to a semi-infinite $su(2)$ representation:
\bea
        J^+_0(J^-_0)^{-n}\ket{n,\mu_2,t}=0,&&\ \ \ \ \mu_1=n\in-\NN\;,\nn
        J^-_0(J^+_0)^{n+1}\ket{n,\mu_2,t}=0,&&\ \ \ \ \mu_1=n\in\NN_0\;.
\eea
The affine extensions are relaxed modules with a charged
singular vector in the extremal diagram. An example is illustrated in
Fig. \ref{charged}.
Observe that one arrow is not double-sided, signaling the presence of
a charged singular vector.

Whenever $\mu_1$ and $\mu_2$ are integers
of different signs, the extremal diagram contains two charged singular
vectors. Neither can be reached from the other by the action of the
zero modes. We shall not be concerned with this case nor with
the cases where $\mu_1$ and $\mu_2$ are integers of the same sign,
as those cases turn out to be irrelevant for our purposes.

\subsection{Singular vectors and embedding diagrams}

Singular vectors in the non-extremal diagram have also been examined
in \cite{FST,SS}. They are essentially obtained by first mapping
an extremal vector in the relaxed module to a (possibly twisted)
highest-weight vector in an auxiliary Verma module. In the latter module,
the singular vectors are known to be (twisted) MFF vectors. When mapped back
to the relaxed module, these `relaxed' MFF vectors satisfy the relaxed
highest-weight conditions. The mappings between the relaxed module
and the auxiliary Verma module will in general be formal and
involve non-integer powers of affine generators. However, the full
construction still makes sense \cite{FST,SS}, in the same way that the
original MFF vectors can be shown to correspond to well-defined vectors.
For certain values of $\mu_1$ and $\mu_2$ the procedure needs extra caution,
but is also treated in \cite{FST,SS}.

\begin{figure}
\centerline{\epsfig{figure=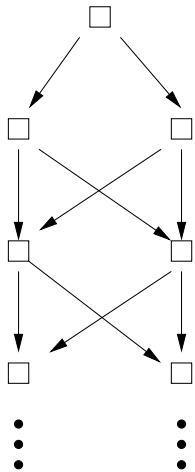,width=2.5cm}}
\vspace{0.3cm}
\caption{Embedding diagram for a relaxed module of type III$_+(0)$.}
\label{Fig.III0}
\end{figure}

\begin{figure}
\centerline{\epsfig{figure=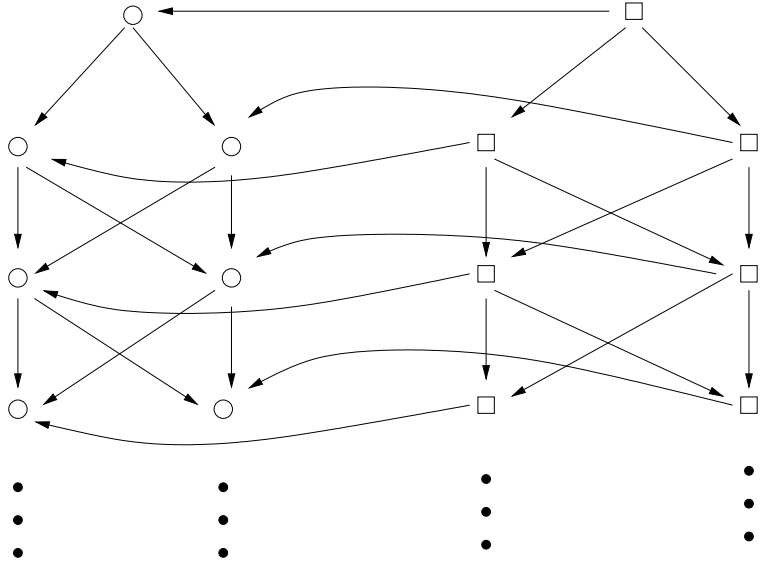,width=10cm}}
\vspace{0.3cm}
\caption{Embedding diagram for a relaxed module of type III$_+(1,-)$.}
\label{Fig.III1-}
\end{figure}

Let us turn to the degeneration patterns of the relaxed modules,
$\RM_{\mu_1,\mu_2,t}$, which will be of interest to us.
In the classification provided
in \cite{FST,SS} (and in the notation of \cite{SS}), they are all of
type III$_+$, meaning that $t\in\QQ_>$, $\mu_1-\mu_2\in\KK(t)$, and
$\mu_1-\mu_2,\ (\mu_1-\mu_2)/t\not\in\ZZ$.
Here we have borrowed the notation \cite{SS}
\ben
        \KK(t)=\{a-bt\ |\ a,b\in\ZZ,\ ab>0\}\;.
\label{K}
\een
There are three interesting sub-cases:\\
$\bullet$ III$_+(0)$: $\mu_1,\mu_2\not\in\ZZ$.\\
The relaxed module has no charged singular vector. Its embedding diagram
is equivalent to the embedding diagram of an ordinary auxiliary Verma
module $\VM_{j,t}$ with spin $j=(\mu_1-\mu_2-1)/2$. This is
illustrated in Fig. \ref{Fig.III0}.
The corresponding submodules appear at the same levels in $\RM$ and $\VM$,
and are nested in identical ways. Relaxed modules of this type are sometimes
referred to as continuous representations.
It is the only type of relaxed module discussed in \cite{Gab}. There it is
denoted ${\cal H}_E$.\\
$\bullet$ III$_+(1,-)$: $\mu_1\in-\NN$.\\
The relaxed module has one charged singular vector. It is
the top open dot in Fig. \ref{Fig.III1-},
and it generates an ordinary Verma module as a submodule of
the relaxed module. Each of the subsequent Verma modules is
embedded via a charged singular vector (illustrated by an open dot)
into the corresponding relaxed module, having a singular vector
(illustrated by an open square) at the same level.\\
$\bullet$ III$_+(1,+)$: $\mu_1\in\NN_0$.\\
The relaxed module has one charged singular vector,
and its embedding diagram is obtained by a mirror reflection
of the one for type III$_+(1,-)$. It should be noted
that the embedded Verma modules are twisted rather than ordinary
Verma modules, with
twist parameter $\theta=1$. Thus, they are simply semi-infinite
lowest-weight Verma modules, while the embedded Verma modules in the type
III$_+(1,-)$ diagrams above are semi-infinite highest-weight Verma modules.

\subsection{Characters of relaxed representations}

The character of the unreduced relaxed module (from which no singular
vectors have been factored out) is easy to write \cite{BFST}:
\ben
       \chi_{\mu_1,\mu_2,t;\theta}^\RM(z,q)
        =\delta(zq^{-\theta},1)\frac{
        q^{\D_{\mu_1,\mu_2,t;\theta}}\ z^{m_{\mu_1,\mu_2,t;\theta}}}{
        \prod_{i\geq 1}(1-q^i)^3}\;,
\label{chiR}
\een
where we have used the notation (\ref{delu1}), i.e.,
$
\delta(u,1) = \sum_{n\in \ZZ} u^n$,
and
the Sugawara dimension and charge of the twisted relaxed
highest-weight state $\ket{\mu_1,\mu_2,t;\theta}$ were introduced in
(\ref{D}). The delta function arises from the boundary of the extremal
diagram (which, in an untwisted case, typically corresponds to a
continuous  representation). Because of it, the character (\ref{chiR}) is
a formal expression which diverges for all $z$. It can be made into a
distribution only on circles $|zq^{-\theta}|=1$ as will be discussed
in the last sections. Nevertheless, the delta function notation refers
to some convenient features. For instance,
one can multiply the expression (\ref{chiR})
by integer powers of $zq^{-\theta}$ without changing its (formal) value.
The superscript $R$ indicates that we deal with  a relaxed module.

Due to the equivalent embedding patterns, characters of irreducible
representations $\RMR$ of type III$_+(0)$ may be obtained in the same
way as characters of the auxiliary irreducible representations $\VMR$
(see Fig. \ref{Fig.III0} and the related discussion).
This means that the character of the irreducible relaxed module may be
written as a delta function times the residue
of a character function associated to an ordinary, irreducible
admissible representation. We will give examples of this in the next section.

For characters originating from the type III$_+(1,-)$ or its
mirror-reflected companion, the
embedding pattern is also similar to the usual one, as illustrated in  Fig.
\ref{Fig.III1-}. The new feature is the regular submodules
generated from the series of charged singular vectors contained in the
relaxed (sub-)modules. The character for an irreducible type
III$_+(1,\pm)$ module may thus formally be obtained by subtracting a regular
character from a delta function. This in turn may be re-written as the
character of a regular representation. This is because the fully reduced
module constructed by factoring out all submodules (including the
one generated by the charged singular vector in the extremal
diagram) is equivalent to a regular module. When the relaxed
module is untwisted, the latter regular module corresponds to an
ordinary highest- or lowest-weight representation, i.e., when the
relaxed module is twisted we have
\bea
       {\rm III}_+(0,+)&:&\ \ \ \RMR_{0,\mu_2,t;\theta}\ \simeq\
                 \VMR_{-(1+\mu_2)/2,t;\theta}\;,\nn
       {\rm III}_+(0,-)&:&\ \ \ \RMR_{-1,\mu_2,t;\theta}\ \simeq\
                 \VMR_{(k-\mu_2)/2,t;\theta+1}\;.
\label{rmr}
\eea
Note that we may freely choose the extremal vector with
$\mu_1=0$ or $\mu_1=-1$, respectively, to characterize the
relaxed module.

\subsection{Almost reduced relaxed modules}

In the main text, we will consider relaxed modules which are not quite
irreducible. They correspond to representations of types
III$_+(1,\pm)$ where the charged
singular vectors are {\em not} removed from the module while all others are.
We shall call them {\em almost reduced} relaxed modules, and denote
them by $\ARM$. Such a module is well-defined, and the resulting character
will turn out to be a delta function times the residue of a regular
representation, i.e., it becomes similar to that of type III$_+(0)$.

The almost reduced relaxed modules may seem ad-hoc, but they appear
naturally in the free-field realization of $\su(2)_{-1/2}$, as we
discuss below. In order to justify further their presence in the conformal
field theory, let us discuss briefly the constraints induced by
`Zhu's algebra'.

\subsection{Relaxed modules and Zhu's algebra}

It is known that by looking at modes of the first non-trivial null vector
of the identity representation, conditions can be obtained on the
fields or representations that are allowed in the theory.
This consideration of `equations of motion' is believed to be
equivalent to studying the formal Zhu's algebra \cite{Zhu}, as described
in \cite{GG}. We have showed before \cite{LMRS} that there
is a null vector, ${\cal N}_4$, at level 4 for $k=-1/2$ in the identity
module. Expanding the null vector in modes
\ben
        {\cal N}_4(z) = \sum_{n\in \ZZ} V_n({\cal N}_4) \ z^{-n-4}\;,
\een
and inserting these modes in three-point functions, we get conditions on
representations that are allowed in the theory.  For example, applying
$V_0({\cal N}_4)$ on a conformal highest-weight state, (i.e.,  one that
is annihilated by $J_{n>0}^a$ in an untwisted module),
we get a condition on the allowed untwisted representations of the form
\ben
        (3+16 C)[3 (J_0^3)^2 - C] = 0\;,
\een
where $C$ is the quadratic Casimir
\ben
       C=J^-_0J^+_0+(J^3_0)^2+J^3_0\;.
\een
Its eigenvalue on the relaxed highest-weight vector is
\ben
       C\ket{\mu_1,\mu_2,t}=\frac{1}{4}\left((\mu_1-\mu_2)^2-1\right)
        \ket{\mu_1,\mu_2,t}\;.
\een
The relaxed modules have an infinite number of $J_0^3$ values and therefore
must have $C=-3/16$ to be allowed in the theory. Thus, the only (untwisted)
relaxed Verma modules allowed are those having Sugawara or conformal
dimension
\ben
       \D_{\mu_1,\mu_2,t}=\frac{C}{k+2}=-\frac{1}{8}\;.
\een
We also see that the difference $\mu_1-\mu_2$ is fixed:
\ben
       |\mu_1-\mu_2|=\hf
\label{mu1mu2}\;.
\een
Similarly, one can derive constraints by acting with other modes, $V_n$,
leading to constraints on the singular vectors in the modules.

Nowhere does the consideration of the `equations of motion' impose
to consider totally reduced relaxed modules. From that perspective,
almost reduced (relaxed) modules are therefore allowed.

\end{document}